\documentclass[twoside]{article}        

\def\minipar{\par\vskip3pt\par}    

\newcount\numsection  
\newcount\numprop  
\newcount\numresult
\def\begprop{\global\advance\numprop by1
\global\numresult=0\medskip\noindent}
\def\result{\global\advance\numresult by1\par\vskip4pt\par\noindent
\hbox{\quad\bf[\arabic{section}.\the\numprop.\the\numresult]\ \ }}
\def\endprop{\medskip\noindent}
\def\prf#1{\medskip\noindent
{\bf Proof of proposition #1$\,$:}\smallskip\noindent}    

\let\epsilon=\varepsilon
\let\phi=\varphi
\usepackage{amssymb}      

\begin{document}
  
\pagenumbering{arabic}
\pagestyle{myheadings} 
\markboth{\noindent\centerline{COECKE AND 
MOORE}}{\noindent\centerline{OPERATIONAL   
GALOIS ADJUNCTIONS}}
\thispagestyle{plain}   
\hbox{}
\par\vskip 2.25 truecm\par    
\noindent{\bf OPERATIONAL GALOIS ADJUNCTIONS}\footnote{Appeared in ``Current Research in
Operational Quantum Logic: Algebras, Categories, Languages'', B.
Coecke, D.J. Moore and A. Wilce (Eds.), Fundamental Theories of
Physics 111, Kluwer Academic Publishers (2000).}  
\par\vskip 0.406 truecm\par
\par\vskip 0.406 truecm
\parindent=2cm\par  
BOB COECKE 
\par {\it Department of Mathematics,     
\par Free University of Brussels,
\par Pleinlaan 2, B-1050 Brussels, Belgium.}  
\par e-mail: bocoecke@vub.ac.be 
\par\vskip 0.406 truecm\par     
DAVID MOORE\footnote{Current address: Department of Physics and
Astronomy, University of Canterbury, PO Box 4800, Christchurch, New
Zealand.} 
\par {\it Department of Theoretical Physics,   
\par University of Geneva,
\par 24 quai Ernest-Ansermet, CH-1211 Geneva 4, Switzerland.
\par}
e-mail: David.Moore@phys.canterbury.ac.nz    
\par\vskip 0.406 truecm\par
\par\vskip 0.406 truecm\par  
\par\vskip 0.406 truecm\par  
\noindent We present a detailed synthetic overview of the utilisation of
categorical techniques in the study of order structures together with
their applications in operational quantum theory. First, after reviewing
the notion of residuation and its implementation at the level of
quantaloids we consider some standard universal constructions and the
extension of adjunctions to weak morphisms. Second, we present the
categorical formulation of closure operators and introduce a hierarchy of
contextual enrichments of the quantaloid of complete join lattices. Third,
we briefly survey physical state-property duality and the 
categorical analysis of derived notions such as causal assignment and the 
propagation of properties.
\parindent=0.6cm
\par\vskip 0.406 truecm\par
\par\vskip 0.406 truecm\par
\noindent {\bf 1. Introduction}  
\numprop=0\setcounter{section}{1}   
\par\vskip 0.406 truecm\par
\par\noindent   
The starting point for the structure theory we shall
expose in this paper is the well known fact that preordered sets may
be considered as small  thin categories; one can then
not only reformulate a  large part of the theory of order structures in
categorical terms, but also  apply general categorical techniques to
specific order theoretic problems.  We provide a brief introduction to
category theory in section 2; for  detailed presentations see for example
[Ad\'amek, Herrlich and Strecker 1990; Borceux 1994; Mac$\,$Lane 1971]. 
First, as discussed in section 3, the notion of an adjunction  reduces to
that of a residuation; the resulting coisomorphy between the categories of
join complete lattices and meet complete lattices will
provide a guiding principle for the rest of this work.  For general
expositions of residuation theory see  [Blyth and Janowitz 1972;
Derd\'erian 1967]. Second, as discussed in section 4, the consideration of
simple examples  allows the direct characterisation of special morphisms
and thereby both  the construction of limits and the definition of
pseudoadjoints for  weak morphisms. Third, as discussed in section 5,
following [Moore 1995, 1997, 2000] the definition of a monad reduces to
that of a closure operator. In  particular, the categories of atomistic
join complete lattices and  closure spaces are equivalent; for a general
discussion of the categorical  algebra of matroids see [Faure 1994; Faure
and Fr\"olicher 1996, 1998]. Fourth, as discussed in section 6, following 
[Amira, Coecke and Stubbe 1998; Coecke and Stubbe 1999a,b, 2000]
the passage from the static consideration of individual lattices to the 
dynamic viewpoint of induced quantaloids allows the introduction of an 
inclusion hierarchy of structures representing successive levels of
contextual enrichment. For general expositions of quantaloid theory
see [Rosenthal
1991].  Fifth, far from being of purely technical
interest, this categorical formalism has direct application in operational
quantum theory, as  developed in [Aerts 1982, 1994; Jauch and Piron 1969;
Piron 1964, 1976, 1990]. In particular, 
as discussed in section 7, following [Moore 1999] the categorical 
equivalence between orthogonal spaces and atomistic complete
ortholattices determined  by  the existence of monadic comparison
functors has a direct interpretation in  terms of the primitive duality
between the state and property descriptions  of a physical system. By
way of application, following [Faure and Fr\"olicher 1993, 1994, 1995]
one can then reformulate the  Hilbertian representation of projective
orthogeometries in purely  categorical terms. Note that our exposition
focuses on the basic structure theory of adjunctions on complete
lattices. As such we shall not discuss topics such as  orthoadjunctions
on orthomodular lattices, an important subject leading to the analysis
of Baer $^\ast$-semigroups via the Sasaki projection [Foulis 1960, 1962]
and the action of conditioning maps on weight spaces [Foulis and Randall
1971, 1974; Frazer, Foulis and Randall 1980]. For extensions to test
spaces see  [Bennett and Foulis 1998; Wilce 2000], and for a logical
analysis of  perfect measurements in this context see  [Coecke and Smets
2000]. 
\par\vglue2.5pt  

For readability and ease of presentation we have relegated proofs 
to an appendix. Note that, while several of our results are either new or
genuine extensions of standard ones, we would like to emphasise our uniform
presentation of the theory and not just its novelty. As such, we have
made no attempt to trace the historical origins and development of our
categorical approach, preferring to emphasise its coherence as a synthetic
tool and its utility in applications. Similarly, our bibliography should be
taken as indicative rather than exhaustive.
For the reader's convenience, we end this introduction by
collecting  together the definitions of the main categories to be
treated in the  following. 
We will use abbreviations when denoting Hom-sets,
e.g., 
$J(L_1,L_2)$ for those of $\hbox{\underbar{JCLatt}}$.
First, any join preserving map $f:L_1\rightarrow L_2$ satisfies
the condition
$f(0_1)=f(\bigvee_1\emptyset)=\bigvee_2f(\emptyset)=
\bigvee_2\emptyset=0_2\,$.
We then obtain the following hierarchy of categories of 
complete lattices$\,$:
%%%%%%%%%%%
%%%%%%%%%%%
%%%%%%%%%%%
%\vfill
%\eject
\par\minipar\minipar\par\noindent
\def\jump{height3pt&\omit&&\omit&&\omit&\cr}
\leftline{\qquad\hbox{\vbox{\offinterlineskip\hrule
\halign{\vrule#&\strut\ \ \ #\hfil\ \ \ &\vrule#&\ \ \ #\hfil\ \ \
&\vrule#&\ \ \ #\hfil\ \ \ &\vrule#\cr height3pt&\omit&&\omit&&\omit&\cr
&Category&&Map preservation&&Constraints&\cr
height3pt&\omit&&\omit&&\omit&\cr
\noalign{\hrule} height3pt&\omit&&\omit&&\omit&\cr 
\jump &WJCLatt&&Non-empty joins&&&\cr
\jump &JCLatt&&Arbitrary joins&&Note$\,$: $f(0_1)=0_2$&\cr
\jump &BJCLatt&&Arbitrary joins, balanced&&$f(1_1)=1_2$&\cr
\jump &DJCLatt&&Arbitrary joins, dense&&$f(a_1)=0_2\Rightarrow
a_1=0_1$&\cr
height3pt&\omit&&\omit&&\omit&\cr}\hrule}}}   

\minipar\noindent 
Dualising we obtain the analogous categories of meet preserving maps. 
Second, the 
Galois adjunction provides an isomorphism between 
\underbar{JCLatt} and \underbar{MCLatt}$^{\rm op}$, which restricts to
isomorphisms 
$\hbox{\underbar{BJCLatt}}\simeq\hbox{\underbar{DMCLatt}}^{\rm op}$ and
$\hbox{\underbar{DJCLatt}}\simeq\hbox{\underbar{BMCLatt}}^{\rm op}$. We
shall show 
that this isomorphism can be extended to weak morphisms in two equivalent
manners. Explicitly, for $L$ a complete lattice let 
$[0,a]=\{\,x\in L\,|\,x<a\}$ and $L^u=L\,\dot{\cup}\,\{{\bf1}\}\,$. Then  
\underbar{WMCLatt}$^{\rm op}$ is isomorphic to each of the categories$\,$:

\par\minipar\minipar\par\noindent 
\def\jump{height3pt&\omit&&\omit&&\omit&\cr}
\leftline{\qquad\hbox{\vbox{\offinterlineskip\hrule
\halign{\vrule#&\strut\ \ \ #\hfil\ \ \ &\vrule#&\ \ \ #\hfil\ \ \
&\vrule#&\ \ \ #\hfil\ \ \ &\vrule#\cr height3pt&\omit&&\omit&&\omit&\cr
&Category&&Hom-sets&&Morphisms&\cr height3pt&\omit&&\omit&&\omit&\cr
\noalign{\hrule} height3pt&\omit&&\omit&&\omit&\cr
&PJCLatt&&$PJ(L_1,L_2)=\cup_{a\in L_1}J(\,[0,a_1],L_2\,)$&&Sectional maps&\cr
\jump  &UJCLatt&&$UJ(L_1,L_2)=BJ(L_1^u,L_2^u)$&&Upper maps&\cr
height3pt&\omit&&\omit&&\omit&\cr}\hrule}}} 

\minipar\noindent
Third, for complete atomistic lattices the duality between 
\underbar{JCLatt} and \underbar{MCLatt} also restricts to atomic
morphisms.
Explicitly, let $\Sigma_{\scriptscriptstyle L}$ be the set of atoms of $L$ 
and $\alpha:\Sigma_1\setminus K_1\rightarrow\Sigma_2$ be a continuous 
partial map between the closure spaces $(\Sigma_1,T_1)$ and
$(\Sigma_2,T_2)\,$.  We then obtain the dual categories$\,$:

\par\minipar\minipar\par\noindent
\def\jump{height3pt&\omit&&\omit&&\omit&\cr}
\leftline{\qquad\hbox{\vbox{\offinterlineskip\hrule
\halign{\vrule#&\strut\ \ \ #\hfil\ \ \ &\vrule#&\ \ \ #\hfil\ \ \
&\vrule#&\ \ \ #\hfil\ \ \ &\vrule#\cr height3pt&\omit&&\omit&&\omit&\cr
&Category&&Morphisms&&Induced from closure&\cr
height3pt&\omit&&\omit&&\omit&\cr
\noalign{\hrule} height3pt&\omit&&\omit&&\omit&\cr
&JCALatt&&$f(\Sigma_{\scriptscriptstyle
L_1})\subseteq\Sigma_{\scriptscriptstyle
L_2}$&&$f_\alpha\!:\!A_1\!\mapsto\!T_2f(A_1\!\setminus\!K_1)$&\cr
\jump &MCALatt&&$(\forall p_1\,\exists
p_2)\,p_1\!<\!g(p_2)$&&$g_\alpha\!:\!A_2\!\mapsto\!K_1\!
\cup\!f^{-1}(A_2)$&\cr
height3pt&\omit&&\omit&&\omit&\cr}\hrule}}} 

\minipar\noindent 
Note that the categories \underbar{JCALatt} and \underbar{CSpace} are 
equivalent. Fourth, applying the power construction to complete
lattices we obtain an inclusion hierarchy of quanta\-loids expressing
successive degrees of contextual  enrichment. Explicitly, define
$P_0(L)=P(L\setminus\{0\})\,$, and  for $f\in J(L_1,L_2)$ and $\theta\in
J(P_0(L_1),P_0(L_2))$ let us write
$f\succ\theta$ if $f(\bigvee A_1)=\bigvee\theta(A_1)$ for each $A_1\in
P_0(L)\,$. We then obtain$\,$:

\par\minipar\minipar\par\noindent
\def\jump{height3pt&\omit&&\omit&&\omit&\cr}
\leftline{\qquad\hbox{\vbox{\offinterlineskip\hrule
\halign{\vrule#&\strut\ \ \ #\hfil\ \ \ &\vrule#&\ \ \ #\hfil\ \ \
&\vrule#&\ \ \ #\hfil\ \ \ &\vrule#\cr height3pt&\omit&&\omit&&\omit&\cr
&Category&&Morphisms&&Name&\cr height3pt&\omit&&\omit&&\omit&\cr
\noalign{\hrule} height3pt&\omit&&\omit&&\omit&\cr &PStruct&&$P_f$ for
$f\in J(L_1,L_2)$&&Power structures&\cr
\jump &BStruct&&$\cup_\alpha P_{f_\alpha}$ for $f_\alpha\in J(L_1,L_2)$&&Based
structures&\cr
\jump &TStruct&&$(f,\theta)$ for $f\succ\theta$&&Transition structures&\cr
\jump &FStruct&&$\theta\in J(P_0(L_1),P_0(L_2))$&&Functional
structures&\cr
height3pt&\omit&&\omit&&\omit&\cr}\hrule}}} 
\par\vskip 0.406 truecm\par
\par\vskip 0.406 truecm\par
\noindent   {\bf 2. Category theory}  
\numprop=0\setcounter{section}{2}   
\par\vskip 0.406 truecm\par
\par\noindent 
At the most naive level, category theory may be construed 
as a hierarchy of object-structure relations, the standard definitions
then reducing to  unicity requirements for induced relations. First, if
the morphism $f$  relates the objects $A$ and $B$ and the morphism $g$
relates the objects
$B$ and $C$ then it is natural to suppose the existence of an induced
relation $g\circ f$ between $A$ and $C\,$: since identification provides a
canonical relation between $A$ and itself we are then led to suppose the
existence of identity morphisms as compositional units; since
$h\circ(g\circ f)$ and $(h\circ g)\circ f$ are both induced relations
between $A$ and $D$ we are led to require associativity considered as
order  indifference of concatenation. We then recover the usual definition
of a category. Second, a relation between morphisms should respect the 
structurally important features involved in the concept of a morphism$\,$:
a functor $F$ should then relate the domain and codomain objects of the
initial morphism to those of the final morphism, so that
$Ff:FA\rightarrow FB$ for $f:A\rightarrow B\,$: identity morphisms form a
distinguished class and so should be preserved; the two induced relations
$F(g\circ f)$ and $Fg\circ Ff$ between $FA$ and $FB$ should coincide. We
then recover the usual definition of a functor. Third, any relation
$\theta$  between two functors $F$ and $G$ should induce a relation or
morphism 
$\theta_{\scriptscriptstyle A}$ between $FA$ and $GA\,$: further for
$f:A\rightarrow B$ the two induced relations 
$\theta_{\scriptscriptstyle B}\circ Ff$ and
$Gf\circ\theta_{\scriptscriptstyle A}$ between $FA$ and $GB$ should
coincide. We then recover the usual definition of a natural
transformation.
\par\vglue2.5pt

Much of the power of category theory is due to the notion of universal
constructions. A local approach is through the dual notions of limits and
colimits of diagrams, where a diagram is an indexed set of objects subject
to structural relations, that is, a functor
$\nabla:\hbox{\underbar{J}}\rightarrow\hbox{\underbar{X}}$ where
\underbar{J} is a small index category encoding the constraints. Now, to
relate diagrams to compatible objects it suffices to remark that  each
object $A$ can be modeled by the corresponding constant functor
$C_{\scriptscriptstyle
A}:\hbox{\underbar{J}}\rightarrow\hbox{\underbar{X}}\,$. By definition, a
source is then a natural transformation
$p:C_{\scriptscriptstyle A}\rightarrow\nabla\,$. In this way we obtain 
limits as distinguished sources $p$ such that for any other source 
$\overline{p}$ there exists a unique morphism $f$ satisfying 
$\overline{p}=p\circ f\,$. On the other hand, a global approach to
universal constructions is provided by the notion of adjoint functors.
Explicitly, let 
$F:\hbox{\underbar{X}}\rightarrow\hbox{\underbar{Y}}$ and 
$G:\hbox{\underbar{Y}}\rightarrow\hbox{\underbar{X}}\,$. Now we can only
directly relate functors with the same domain and codomain. We are then
led to define $F\dashv G$ if there exist  natural transformations 
$\eta:{\rm Id}_{\scriptscriptstyle\underline{\rm X}}\rightarrow G\circ F$ 
and
$\epsilon:F\circ G\rightarrow{\rm Id}_{\scriptscriptstyle\underline{\rm
Y}}$  satisfying the coherence conditions
$\epsilon F\circ F\eta={\rm id}F$ and $G\epsilon\circ\eta G={\rm id}G\,$.
Note that the two conceptions of universality are closely interrelated.
For
example, if $F\dashv G$ then $G$ preserves limits and $F$ preserves
colimits, and any two (co)adjoints of the same functor are naturally
isomorphic. Further, products, defined as limits of trivial diagrams whose
index  category has only identity morphisms, may be globalised as adjoints
of the appropriate diagonal functors. Finally, two categories are called
equivalent if there exist natural isomorphisms 
$\phi:{\rm Id}_{\scriptscriptstyle\underline{\rm X}}\rightarrow G\circ F$
and 
$\psi:{\rm Id}_{\scriptscriptstyle\underline{\rm Y}}\rightarrow F\circ G$
satisfying the equivalent coherence conditions $F\phi=\psi F$ and
$\phi G=G\psi\,$. Note that in this case $F\dashv G\dashv F\,$,
the two functors $F$ and $G$ then preserving (co)limits and adjunctions.
\vfill
\eject 
Often one is interested in categories \underbar{A} which can be
considered  as specialisations of some base category \underbar{X}$\,$.
For example, we  may wish to treat objects in \underbar{A} as objects in
\underbar{X} together with extra structure, and morphisms in
\underbar{A} as morphisms  in \underbar{X} which respect that structure
in some sense. We are then led to define a concrete category over the
category \underbar{X} to be a  pair $(\hbox{\underbar{A}},U)\,$, where
\underbar{A} is a category  and
$U$ is a faithful functor from \underbar{A} to \underbar{X}$\,$,  that is,
a functor which is injective on Hom-sets. It is then of interest  to
consider those morphisms $f:UA\rightarrow UB$ in \underbar{X} which  lift,
in the sense that there exists $\phi:A\rightarrow B$ with
$U\phi=f\,$. For example, $\phi:A\rightarrow B$ is called initial if
$f:UA'\rightarrow UA$ lifts whenever $U\phi\circ f$ does, or final if
$g:UB\rightarrow UB'$ lifts whenever $f\circ U\phi$ does. Next, each fibre
$F(X)=U^{-1}(X)$ has a natural preorder defined by $A\prec B$ if 
${\rm id}_{\scriptscriptstyle X}\!:\!UA\!\rightarrow\!UB$ lifts. In 
particular, categories whose fibres are ordered by equality have algebraic 
character, whereas those whose fibres are complete lattices have 
topological character. Finally, an important step in category theory is to
replace Hom-sets by objects in an appropriate structure category,
leading  to the notion of enrichment. Explicitly, in order to define
composition adequately we require that the structure category
\underbar{V} be symmetric
monoidal closed, in the sense that there exists a tensor functor
$\otimes:\hbox{\underbar{V}}\times\hbox{\underbar{V}}\rightarrow
\hbox{\underbar{V}}$
and unit object $I\in{\rm Ob}(\,\hbox{\underbar{V}}\,)$ such that the
usual
composition laws may be replaced by coherent natural transformations. For 
example, standard categories are just categories over \underbar{Set}$\,$, 
$2$-categories are defined to be categories over \underbar{Cat}$\,$, and 
quantaloids may be considered as categories enriched in join complete
lattices [Borceux and Stubbe 2000].  
Quantales then represent the monoidal one-object restrictions
[Paseka and Rosick\'y 2000].  We obtain quantale
and quantaloid morphisms as the corresponding
$\underbar{JCLatt}$-enriched functors.    
\par\vglue2.5pt

Finally, much of the theory of order structures can be reformulated in 
purely categorical terms by remarking that preordered classes are in 
bijective correspondence with thin categories, namely categories for which 
each Hom-set has at most one element. Explicitly, $a<b$ if and only if
${\rm Hom}(a,b)\not=\emptyset\,$, reflexivity being the   existence of
identity morphisms and transitivity being the condition of  morphism
composability. Note that the unicity of a morphism 
$\alpha:a\rightarrow b$ implies that any diagram which can be written must 
commute. In particular, products are exactly meets,
$(\forall\alpha\in\Omega)\,(x<a_\alpha)\Leftrightarrow x<\wedge_\alpha
a_\alpha\,$, whereas coproducts are exactly joins,
$(\forall\alpha\in\Omega)\,(a_\alpha<x)\Leftrightarrow\vee_\alpha
a_\alpha<x\,$. Further, a functor between two preordered classes is
exactly an isotone map, preservation of order being exactly preservation
of composition. In particular, there exists a natural transformation
$\theta:f\rightarrow g$ if  and only if $f<g\,$, this being the necessary
and sufficient  condition for the existence of morphisms 
$\theta_a:f(a)\rightarrow g(a)\,$. For isotone maps $f:L_1\rightarrow L_2$
and $g:L_2\rightarrow L_1\,$,  we then have that $f\dashv g$ if and only
if ${\rm id}_1<(g\circ f)$ and 
$(f\circ g)<{\rm id}_2\,$, these being the necessary and sufficient 
conditions for the existence of natural transformations
$\eta:{\rm id}_1\rightarrow(g\circ f)$ and
$\epsilon:(f\circ g)\rightarrow{\rm id}_2\,$. Note that in this context
quantaloids are exactly locally complete and thin 
$2$-categories. For convenience, in the following we shall restrict our 
attention to posets, that is, small thin categories for which no two 
distinct elements are isomorphic. For such categories natural isomorphy 
reduces to identity, so that adjoints are unique when they exist. 
\vfill
\eject
\par\vskip 0.406 truecm\par
\par\vskip 0.406 truecm\par
\noindent  {\bf 3. Morphisms and adjunctions}  
\numprop=0\setcounter{section}{3}   
\par\vskip 0.406 truecm\par
\par\noindent    
In this preliminary section we recall some elementary
facts about  adjunctions on posets considered as thin categories, results
which form  the core of the rest of this work. Explicitly, we start by
transcribing the definition and basic properties of adjunctions in the
context of posets, before turning to limit preservation properties in the
context of complete lattices. We then globalise these observations to
categories of posets and finish with some remarks on the orthocomplemented
case. Let $f$ and $g$ be isotone maps on posets. Then$\,$:
\begprop
\result ${\rm id}_1<(g\circ f)$ {\&} $(f\circ g)<{\rm id}_2$ iff
$f(a_1)<a_2\Leftrightarrow a_1<g(a_2)\,$;
\result $f$ is an isomorphism with inverse $g$ iff $f\dashv g\dashv f\,$;
\result If $f\dashv g$ then $f\circ g\circ f=f$ and $g\circ f\circ g=g\,$;
\result If $f\dashv g$ and $\overline{f}\dashv\overline{g}$ then
$f<\overline{f}\Leftrightarrow\overline{g}<g\,$;
\result ${\rm id}\dashv{\rm id}\,$, and if $f\dashv g$ and
$\overline{f}\dashv\overline{g}$ then $(\,\overline{f}\circ
f)\dashv(g\circ\overline{g}\,)\,$.
\endprop
\par\noindent  
The first result gives a more practical form for the adjunction
condition, whereas the second encodes isomorphy as equivalence. The third
result  establishes adjunctions as pseudoinverses. As we shall see later,
the  fourth result enables a globalisation of adjunctions from the level
of  individual posets to the level of categories of posets, the fifth 
leading to a natural generalisation to quantaloids. Next, transcribing 
the limit preservation properties of adjunctions we obtain the following 
results for isotone maps on complete lattices$\,$:
\begprop
\result If $f\dashv g$ then $f(\bigvee A_1)=\bigvee f(A_1)$ and
$g(\bigwedge A_2)=\bigwedge g(A_2)\,$;
\result If $f(\bigvee\!A_1)\!=\!\bigvee\!f(A_1)$ then 
$f\dashv f^\ast\!\!:\!L_2\!\rightarrow\!L_1\!:\!a_2\!\mapsto\!\!
\bigvee\{a_1\!\in\!L_1\,|\,f(a_1)\!<\!a_2\}\,$;
\result If $g(\bigwedge A_2)\!=\!\bigwedge g(A_2)$ then 
$g\vdash g_\ast\!:\!L_1\!\rightarrow\!L_2\!:\!a_1\!\mapsto\!
\bigwedge\{a_2\!\in\!L_2\,|\,a_1\!<\!g(a_2)\}\,$;
\result If $f_\alpha\dashv g_\alpha$ then 
$\vee_\alpha f_\alpha\dashv\wedge_\alpha\,g_\alpha\,$; 
\result If $f\dashv g\,$, $f_\alpha\dashv g_\alpha$ then 
$f\circ(\vee_\alpha f_\alpha)\dashv\wedge_\alpha(g_\alpha\circ g)\,$, 
$(\vee_\alpha f_\alpha)\circ f\dashv\wedge_\alpha(g\circ g_\alpha)\,$.
%\endprop
\par\medskip\noindent
The first result implies that join (meet) preservation is a
necessary  condition for the existence of a right (left) adjoints, the
second and third implying sufficiency. The fourth result implies that the 
set $J(L_1,L_2)$ of join preserving maps between the complete lattices 
$L_1$ and $L_2$ is a complete lattice with respect to the pointwise join,
whereas the set
$M(L_2,L_1)$ of meet preserving maps between the complete lattices
$L_2$ and $L_1$ is a complete lattice with respect to the pointwise meet.
Finally, the fifth result implies that composition distributes on both 
sides over joins in \underbar{JCLatt} and meets in \underbar{MCLatt}$\,$. 
In particular, the category \underbar{JCLatt} provides the paradigm example
of a quantaloid.

\minipar

In our last remarks we have implicitly used Birkhoff's theorem, which
states  that any join complete lattice is also meet complete and
conversely. Note  that this result may be construed as an application of
the adjoint functor  theorem to the indexed diagonal functor 
$\Delta:L\rightarrow\times_\alpha L:a\mapsto(a_\alpha=a)\,$, since $L$ is
complete if and only if $J\dashv\Delta\dashv M\,$, with $J$  the join and
$M$ the meet. Nevertheless, as is typical in category theory,  join and
meet completeness are rather different at the level of morphisms.  For
example, let $M(L_1,L_2)^{\rm co}$ be the complete lattice of meet 
preserving maps with opposite pointwise order, \underbar{MCLatt}${}^{\rm
op}$  be the category of complete lattices with meet preserving maps and
opposite  Hom-sets, and \underbar{MCLatt}${}^{\rm coop}$ be the quantaloid of
complete  lattices with meet preserving maps, opposite pointwise order and
opposite  Hom-sets. Considering the families of maps 
$A^\ast\!:\!L\!\mapsto\!L;\,f\!\mapsto\!f^\ast$ and 
$A_\ast\!:\!L\!\mapsto\!L;\,g\!\mapsto\!g_\ast$ we then obtain$\,$:
\begprop
\result $J(L_1,L_2)$ is isomorphic as a complete lattice to
$M(L_2,L_1)^{\rm co}$;
\result \underbar{JCLatt} is isomorphic as a category to
\underbar{MCLatt}${}^{\rm op}$;
\result \underbar{JCLatt} is isomorphic as a quantaloid to
\underbar{MCLatt}${}^{\rm coop}$.
\endprop
\par\noindent  Finally, the above dualities can be usefully restricted to
orthocomplemented lattices, that is, bounded lattices $L$ 
equipped with an
operation
$':L\rightarrow L$ satisfying$\,$: $a<b\Rightarrow b'<a'$; $a''=a\,$;
$a\wedge a'=0\,$. Explicitly, for $\alpha:L_1\rightarrow L_2$ a map
between orthocomplemented  lattices let 
$C(\alpha):L_1\rightarrow L_2:a_1\mapsto\alpha(a_1')'$
be the conjugate map. For $f\dashv g$ we then define the orthoadjoints 
$f^\dagger:L_2\rightarrow L_1:a_2\mapsto g(a_2')'$ and
$g_\dagger:L_1\rightarrow L_2:a_1\mapsto f(a_1')'$. Let us write
\underbar{IoLatt} for the category of orthocomplemented  lattices with
isotone maps, \underbar{JCoLatt} and \underbar{MCoLatt} for  the
categories of complete orthocomplemented lattices with respectively  join
or meet preserving maps, and \underbar{COLatt} for the category of 
complete lattices with maps preserving the join, the meet, and the 
orthocomplementation. Then$\,$:
\begprop  
\result $C$ is an endofunctor on \underbar{IoLatt} restricting to
$\hbox{\underbar{JCoLatt}}\simeq\hbox{{MCoLatt}}\,$; 
\result In \underbar{JCoLatt} we have
$(f_1\!\circ\!f_2)^\dagger\!=\!f_2^\dagger\!\circ\!f_1^\dagger$,
$f^{\dagger\dagger}\!=\!f$,
$f^\dagger\!\circ\!f\!=\!0_1\Leftrightarrow f\!=\!0_2\,$;
\result In \underbar{JCoLatt} we have $u^\dagger\circ u={\rm id}$ iff
$a_1<b_1'\Leftrightarrow u(a_1)<u(b_1)'$;
\result In \underbar{COLatt} we have $h_\ast(a_2')=h^\ast(a_2)'$,
$h^\dagger=h_\ast\,$, $h\circ h^\dagger\circ h=h\,$.
\endprop
\par\noindent The first result implies that \underbar{JCoLatt} is
self-dual. The second  result exhibits \underbar{JCoLatt} as a regular
$\dagger$-semigroup. The third result classifies
isometries in \underbar{JCoLatt}$\,$, whereas the fourth implies
that all orthomorphisms are partially isometric.
\par\vskip 0.406 truecm\par
\par\vskip 0.406 truecm\par
%%%%%%%%%%
%\vfill
%\eject
%%%%%%%%%%
\noindent   {\bf 4. Special morphisms}  
\numprop=0\setcounter{section}{4}   
\par\vskip 0.406 truecm\par
\par\noindent    As is usual in category theory, it is important to have a
number of  examples of particular morphisms such as constant maps or
subobject  inclusions.  First, for ${\bf2}$ the two-element lattice and 
$[0,a]=\{\,x\in L\,|\,x<a\,\}$ a lower interval, let
\par\medskip
\par\noindent\leftline{\hbox{\vbox{\baselineskip=15pt\halign{
\qquad$#\,$&$\,#\,$\qquad\hfil&$#\,$&$\,#\,$\hfil\cr
\alpha_a:&{\bf 2}\rightarrow L:0\mapsto0\,;\ 1\mapsto a
&C^a:&L\rightarrow{\bf 2}:x\mapsto1\,(a<x)\,;\ 0\,(a\not<x)\,;\cr
\alpha^a:&{\bf 2}\rightarrow L:0\mapsto a\,;\ 1\mapsto1
&C_a:&L\rightarrow{\bf 2}:x\mapsto0\,(x<a)\,;\ 1\,(x\not<a)\,;\cr
&&&\cr
i_a:&[0,a]\rightarrow L:x\mapsto x
&\hat{\imath}{}_a:&[0,a]\rightarrow L:x\mapsto x\,(x\not=a)\,;\ 1\,(x=a)\,;\cr
\pi_a:&L\rightarrow[0,a]:x\mapsto x\wedge a
&\hat{\pi}{}_a:&L\rightarrow[0,a]:x\mapsto x\,(x<a)\,;\ a\,(x\not<a)\,.\cr
}}}}\par\medskip\par\noindent
For any adjunction $f\dashv g$ we then
obtain$\,$:
\begprop
\result $\alpha_a\dashv C^a$ with $f\circ\alpha_{a_1}=\alpha_{f(a_1)}$ and
$C^{a_1}\circ g=C^{f(a_1)}\,$;
\result $C_a\dashv\alpha^a$ with $\,g\circ\alpha^{a_2}=\alpha^{g(a_2)}$
and $\,C_{a_2}\circ f=C_{g(a_2)}\,$;
\result $i_a\dashv\pi_a$ with $\pi_a\circ i_a={\rm id}\,$, and
$\hat{\pi}{}_a\dashv\hat{\imath}{}_a$ with
$\hat{\pi}{}_a\circ\hat{\imath}{}_a={\rm id}\,$.
\endprop
\par\noindent  Note that the $\alpha_a$ exhaust join preserving maps 
$\alpha:{\bf2}\rightarrow L\,$, since any such map must satisfy 
$\alpha(0)=0\,$, whereas the $\alpha^a$ exhaust meet  preserving maps
$\alpha:{\bf2}\rightarrow L\,$, since any such map  must satisfy
$\alpha(1)=1\,$. In particular, the $C^a$  exhaust meet preserving maps
$C:L\rightarrow{\bf2}$ whereas the $C_a$  exhaust join preserving maps
$C:L\rightarrow{\bf2}\,$. Here the restriction to complete join maps is
essential; indeed the kernels of finite join maps
$f:L\rightarrow{\bf2}$ are exactly the ideals on $L\,$, such an ideal
being  prime if and only if $f$ also preserves finite meets. Now, if
$f:L_1\rightarrow L_2$ preserves joins then $f(0_1)=0_2\,$:
we then call $f$ balanced if $f(1_1)\!=\!1_2$ or dense if 
$f(a_1)\!=\!0_2\Leftrightarrow a_1\!=\!0_1$. Dually, if 
$g:L_2\rightarrow L_1$ preserves meets then $g(1_2)=1_1\,$:
we then call $g$ balanced if $g(0_2)\!=\!0_1$ or dense if
$g(a_2)\!=\!1_1\Leftrightarrow a_2\!=\!1_2$. Note that a morphism in
either \underbar{JCLatt} or \underbar{MCLatt} is injective only  if it 
is dense, or surjective only if it is balanced. Recall that 
$h$ is called an epimorphism if
$h_1\circ h=h_2\circ h\Rightarrow h_1=h_2\,$, or a monomorphism if 
$h\circ h_1=h\circ h_2\Rightarrow h_1=h_2\,$. For $f\dashv g$ an adjunction
we then obtain$\,$:
\begprop
\result $f$ is balanced iff $g$ is dense$\,$, or dense iff $g$ is
balanced$\,$;
\result $f$ is epic iff $f$ is surjective iff $f\circ g={\rm id}_2$ iff
$g$ is injective iff $g$ is monic$\,$;
\result $f$ is monic iff $f$ is injective iff $g\circ f={\rm id}_1$ iff
$g$ is surjective iff $g$ is epic$\,$.
\endprop
\par\noindent The first result implies that the categories
\underbar{BJCLatt} with  balanced join morphisms and
\underbar{DMCLatt}${}^{\rm op}$ with opposite dense meet morphisms are
isomorphic, as are the categories 
\underbar{DJCLatt} and \underbar{BMCLatt}${}^{\rm op}$. The second and 
third results render explicit the general duality between
epimorphisms and monomorphisms together with their standard set 
theoretic interpretations. Finally, if 
$\overline{f}\circ f={\rm id}$ then $f$ is called a 
section and $\overline{f}$ is called a retraction; each 
section is monic and each retraction is epic. For example,
$(\hat{\pi}{}_a\circ i_a)(x)=\hat{\pi}{}_a(x)=x$, so that
$\hat{\pi}{}_a$ is a retraction and $i_a$ is a section in 
\underbar{JCLatt}$\,$. Similarly,
$(\pi_a\circ\hat{\imath}{}_a)(x)
=[\pi_a(x)\,(x\!\not=\!a);\,\pi_a(1)\,(x\!=\!a)]=x$, so that 
$\pi_a$ is a retraction and $\hat{\imath}{}_a$ is a section in
\underbar{MCLatt}$\,$.

\minipar

Next, recall that the direct product $\times_\alpha L_\alpha$ of the
family of bounded posets $L_\alpha$ is the Cartesian product of the 
$L_\alpha$ equipped with the pointwise order. 
We may then define the maps$\,$:
\par\medskip
\par\noindent\leftline{\hbox{\vbox{\baselineskip=15pt\halign{
\qquad$#\,$&$\,#\,$\hfil\cr
\Pi_\beta:&\times_\alpha L_\alpha\rightarrow L_\beta
:(a_\alpha)\mapsto a_\beta\,;\cr
i_\beta:&L_\beta\rightarrow\times_\alpha L_\alpha:b\mapsto(a_\alpha)\ 
\hbox{with}\  a_\alpha=b\ (\alpha=\beta)\,;\ 0\ (\alpha\not=\beta)\,;\cr
j_\beta:&L_\beta\rightarrow\times_\alpha L_\alpha:b\mapsto(a_\alpha)\ 
\hbox{with}\ a_\alpha=b\ (\alpha=\beta)\,;\ 1\ (\alpha\not=\beta)\,.\cr
}}}}\par\medskip\par\noindent 
On the other hand, recall that the horizontal
sum $\dot{\cup}_\alpha L_\alpha$ of the $L_\alpha$ is the disjoint union of
the $L_\alpha\setminus\{0_\alpha,1_\alpha\}$ with componentwise order and 
adjoined minimal and maximal elements. 
We may then define the maps$\,$:
\par\medskip
\par\noindent\leftline{\hbox{\vbox{\baselineskip=15pt\halign{
\qquad$#\,$&$\,#\,$\hfil\cr
{\rm I}_\beta:&L_\beta\rightarrow\dot{\cup}{}_\alpha L_\alpha
:b\mapsto b\,;\cr
\sigma_\beta:&\dot{\cup}{}_\alpha L_\alpha\rightarrow L_\beta
:x\mapsto x\ (x\in L_\beta)\,;\ 1\ (x\notin L_\beta)\,;\cr
\rho_\beta:&\dot{\cup}{}_\alpha L_\alpha\rightarrow L_\beta
:x\mapsto x\ (x\in L_\beta)\,;\ 0\ (x\notin L_\beta)\,.\cr
}}}}\par\medskip\par\noindent 
Finally, recall that a product of the family of objects
$L_\beta$ is an object $\overline{L}$  together with a family of morphisms
$\overline{p}{}_\beta:\overline{L}\rightarrow L_\beta$ such that for any
object $L$ and any family of morphisms 
$p_\beta:L\rightarrow L_\beta$ there exists a unique morphism 
$\theta:L\rightarrow\overline{L}$ satisfying
$p_\beta=\overline{p}{}_\beta\circ\theta\,$; coproducts being
products in the opposite category. Then$\,$:  
\begprop
\result $i_\beta\dashv\Pi_\beta\dashv j_\beta$ and 
$\sigma_\beta\dashv{\rm I}_\beta\dashv\rho_\beta\,$;
\result $\Pi_\beta$ is the product in \underbar{BPos}$\,$,
\underbar{JCLatt} and \underbar{MCLatt}$\,$;
\result The coproducts in \underbar{BPos}$\,$, \underbar{JCLatt}$\,$,
\underbar{MCLatt} are ${\rm I}_\beta\,$, $i_\beta\,$,
$j_\beta\,$.
\endprop
\par\noindent 
In particular, $\Pi_\beta$ preserves both the join and the meet,
$i_\beta$  preserves the join and $j_\beta$ preserves the meet. Further,
$i_\beta$ preserves non-empty meets however $i_\beta(1)\not=1\,$, whereas
$j_\beta$ preserves non-empty joins however $j_\beta(0)\not=0\,$. We then
have simple  examples of weak morphisms. On the other hand, ${\rm
I}_\beta$ preserves both the join and the meet, 
$\sigma_\beta$ preserves the join and $\rho_\beta$ preserves the meet. 
Further, $\sigma_\beta(1)=1$ however $\sigma_\beta$ does not preserve 
binary meets, whereas $\rho_\beta(0)=0$ however $\rho_\beta$ does not 
preserve binary joins. We then have simple examples of balanced morphisms.
 
\minipar

Finally, in applications one is often led to consider maps 
$f:L_1\rightarrow L_2$ which preserve non-empty joins or maps 
$g:L_2\rightarrow L_1$ which preserve non-empty meets. In this way we
obtain the categories \underbar{WJCLatt} and \underbar{WMCLatt}$\,$. While
such maps do not possess strict Galois adjoints, by  considering either
codomain restrictions or pointed extensions we  can nevertheless introduce
the notion of weak adjunctions. Explicitly, let $g:L_2\rightarrow L_1$
preserve non-empty meets but not  necessarily satisfy $g(1_2)=1_1\,$. Then,
for $L^u=L\,\dot{\cup}\,\{{\bf1}\}$ the upper pointed extension with 
adjoined universal maximal element $a<{\bf1}\,$, the two maps
$g^p:L_2\!\rightarrow\![0_1,g(1_2)]:a_2\!\mapsto\!g(a_2)$ and
$g^u:L_2^u\!\rightarrow\!L_1^u:a_2\!\mapsto\!g(a_2);\,{\bf1}\!\mapsto\!{\bf1}$ 
do preserve arbitrary meets, and so have respective left adjoints
$f^p:[0_1,g(1_2)]\rightarrow L_2$ and $f^u:L_1^u\rightarrow L_2^u$. Now
$f^p$ is of the form $\alpha_1:[0_1,a_1]\!\rightarrow\!L_2$ whereas $g^u$
is dense, $g^u(a_2)\!=\!{\bf1}\Leftrightarrow a_2\!=\!{\bf1}\,$, so that 
$f^u$ is balanced, $f^u({\bf1})\!=\!{\bf1}$. We are then led to consider the
categories \underbar{PJCLatt} with sectional partial morphisms,
$P\!J(L_1,L_2)=\cup_{a_1\in L_1}J(\,[0_1,a_1]\,,L_2\,)\,$, and
\underbar{UJCLatt} with upper pointed morphisms,
$U\!J(L_1,L_2)=B\!J(\,L_1^u\,,L_2^u\,)\,$. Note that for
maps $\alpha_1:[0_1,a_1]\rightarrow L_2$ and 
$\alpha_2:[0_2,a_2]\rightarrow L_3$ we have
$(\alpha_2\circ\alpha_1):[0_1,\alpha_1^\ast(a_2)]\rightarrow L_3\,$, since
$\alpha_1(x_1)\in[0_2,a_2]\Leftrightarrow\alpha_1(x_1)<a_2\Leftrightarrow
x_1<\alpha_1^\ast(a_2)\,$. Now, given morphisms $\alpha\in
P\!J(L_1,L_2)$ and $F\in U\!J(L_1,L_2)$  let us define the maps
\par\medskip
\par\noindent\leftline{\hbox{\vbox{\baselineskip=15pt\halign{
\qquad$#\,$&$\,#\,$\hfil\cr
F_\alpha:&L_1^u\!\rightarrow\!L_2^u:x_1\!\mapsto
\!\alpha(a_1)\,(x_1\!<\!a_1)\,;\
{\bf1}\,(x_1\!\not<\!a_1)\,;\cr
G_\alpha:&L_2^u\!\rightarrow\!L_1^u:x_2\!\mapsto\!\alpha^\ast(x_2)\,;\
{\bf1}\!\mapsto\!{\bf1}\,;\cr
&\cr
\alpha_{\scriptscriptstyle F}:&[0_1,F^\ast(1_2)]\rightarrow L_2
:x_1\mapsto F(x_1)\,;\cr
\beta_{\scriptscriptstyle F}:&L_2\rightarrow[0_1,F^\ast(1_2)]
:x_2\mapsto F^\ast(x_2)\,.\cr }}}}\par\medskip\par\noindent 
For $g\in W\!M(L_1,L_2)$ we then obtain$\,$:
\begprop
\result $F_\alpha\dashv G_\alpha$ with
$F_{\alpha_2\circ\alpha_1}=F_{\alpha_2}\circ F_{\alpha_1}$ and
$F_{\alpha_F}=F\,$;
\result $\alpha_{\scriptscriptstyle F}\dashv\beta_{\scriptscriptstyle F}$
with $\alpha_{\scriptscriptstyle F_2\circ F_1}=\alpha_{\scriptscriptstyle
F_2}\circ\alpha_{\scriptscriptstyle F_1}$ and $\alpha_{\scriptscriptstyle
F_\alpha}=\alpha\,$;
\result If $\alpha\dashv g^p$ and $F\dashv g^u$ then
$\alpha=\alpha_{\scriptscriptstyle F}$ and $F=F_\alpha\,$.
\endprop
\par\noindent 
The first and second results enable us to pass from partial
morphisms to pointed morphisms and conversely in a functorial manner, the
third result implying that the two methods of defining weak adjoints are
equivalent. In particular, the three categories \underbar{PJCLatt}$\,$, 
\underbar{WMCLatt}$^{\rm op}$ and \underbar{UJCLatt} are isomorphic.
\par\vskip 0.406 truecm\par
\par\vskip 0.406 truecm\par
\noindent   {\bf 5. Monads and closure operators}  
\numprop=0\setcounter{section}{5}   
\par\vskip 0.406 truecm\par
\par\noindent    Recall that a monad on the category \underbar{X} is a
triple $(T,\eta,\mu)$  consisting of an endofunctor $T$ together with
natural transformations 
$\eta:{\rm Id}\rightarrow T$ and $\mu:T\circ T\rightarrow T$ satisfying
the
coherence conditions$\,$: 
$\mu\circ T\mu=\mu\circ\mu T\,$; $\mu\circ T\eta={\rm id}T\,$;
$\mu\circ\eta T={\rm id}T\,$. Note that if $L\dashv R$ is an adjunction
via the natural transformations
$\eta$ and $\epsilon$ then $(R\circ L,\eta,R\epsilon L)$ is a monad. On
the
other hand, if $(T,\eta,\mu)$ is a monad then setting
${\rm Ob}(\hbox{\underbar{X}}^{\scriptscriptstyle T})$ to be the set  of
$T$-algebras $(A,\alpha)\,$, where $\alpha:TA\rightarrow A$ satisfies
$\alpha\circ\eta_{\scriptscriptstyle A}={\rm id}_{\scriptscriptstyle A}$
and $\alpha\circ T\alpha=\alpha\circ\mu_{\scriptscriptstyle A}\,$, and
${\rm Hom}((A,\alpha),(B,\beta))$ to be the set of $T$-morphisms
$f:A\rightarrow B\,$, where $f\circ\alpha=\beta\circ Tf\,$, we obtain the
so-called Eilenberg-Moore category generated by the adjunction
$F^{\scriptscriptstyle T}\dashv U^{\scriptscriptstyle T}$, where
$F^{\scriptscriptstyle T}:A\mapsto(TA,\mu_{\scriptscriptstyle A})\,;\
f\mapsto Tf$ and
$U^{\scriptscriptstyle T}:\,(A,\alpha)\mapsto A\,;\ f\mapsto f\,$.
Further, for any monad of the form $T=R\circ L$ there exists a unique
functor
$K:\hbox{\underbar{Y}}\rightarrow\hbox{\underbar{X}}^{\scriptscriptstyle
T}$ such that $R=U^{\scriptscriptstyle T}\circ K$ and 
$F^{\scriptscriptstyle T}=K\circ L\,$. Explicitly,
$K:A\mapsto(RA,R\epsilon_{\scriptscriptstyle A})\,;\ f\mapsto Rf\,$. Now
in the context of posets considered as thin categories the coherence 
conditions are trivially satisfied, since any diagram which can be written
must commute. An isotone map $T:L\rightarrow L$ is then a monad if and
only if  the natural transformations $\eta$ and $\mu$ exist, which is the
case if and only if ${\rm id}<T$ and $T\circ T<T$ so that $T$ is a closure
operator. First, there exists a morphism $\alpha:Ta\rightarrow a$ if and
only if
$Ta<a\,$, which is the case if and only if $Ta=a\,$. The category 
$L^{\scriptscriptstyle T}$ is then the set of fixed points of $T\,$. 
Second, since $F^{\scriptscriptstyle T}\dashv U^{\scriptscriptstyle T}$ we
have that $F^{\scriptscriptstyle T}$ preserves colimits and 
$U^{\scriptscriptstyle T}$ preserves limits. In particular, if $L$ is a
complete lattice then so is $L^{\scriptscriptstyle T}$, with
$\bigwedge_{\scriptscriptstyle T}A=\bigwedge A$ and 
$\bigvee_{\scriptscriptstyle T}A=T(\bigvee A)\,$. Third, if $f\dashv g$
then $f\circ g\circ f=f$ and $g\circ f\circ g=g\,$, so that the 
Eilenberg-Moore category of the monad $g\circ f$ is given by 
the image of $g\,$.

\minipar

Recall that an atom of the lattice $L$ is a minimal nonzero element and 
that the complete lattice $L$ is called atomistic if 
$a=\bigvee\{\,p\in\Sigma_{\scriptscriptstyle L}\,|\,p<a\,\}$ for each 
$a\in L\,$, where $\Sigma_{\scriptscriptstyle L}$  is the set of atoms of
$L\,$. A closure operator $T:L\rightarrow L$ on the atomistic lattice is
then called simple if $T(0)=0$ and $T(p)=p$ for  each
$p\in\Sigma_{\scriptscriptstyle L}\,$. Since the atoms are all fixed
points, the Eilenberg-Moore category associated to any simple closure 
operator is also atomistic. As we shall now
indicate, we then obtain an equivalence between suitable categories of
complete atomistic lattices and closure spaces, namely sets $\Sigma$
equipped with a simple closure operator on $P(\Sigma)\,$. Explicitly, let
$f\dashv g$ be an adjunction between complete atomistic  lattices and
$\alpha:\Sigma_1\setminus K_1\rightarrow\Sigma_2$ be a partially defined
map between closure spaces. Then$\,$:
\begprop
\result $f(\Sigma_{\scriptscriptstyle
L_1})\subseteq\Sigma_{\scriptscriptstyle L_2}\cup\{0_2\}$ iff $(\forall
p_1\in\Sigma_{\scriptscriptstyle L_1})(\exists
p_2\in\Sigma_{\scriptscriptstyle L_2})\,p_1<g(p_2)\,$;
\result $\alpha(T_1A_1\setminus K_1)\subseteq T_2\alpha(A_1\setminus K_1)$
iff $K_1\cup\alpha^{-1}(A_2)$ is closed for $A_2$ closed;
\result The above conditions are stable under composition.
\endprop
\par\noindent 
We then obtain the dual categories \underbar{JCALatt} and
\underbar{MCALatt} together with the category
\underbar{CSpace} of  closure spaces. Next, given morphisms
$\alpha\in C\!S((\Sigma_1,T_1),(\Sigma_2,T_2))$ 
and
$f\in J\!A(L_1,L_2)$ 
let us define
\par\medskip
\par\noindent\leftline{\hbox{\vbox{\baselineskip=15pt\halign{
\qquad$#\,$&$\,:#\,$\hfil\ &\ $#\,$&$\,:#\,$\hfil\cr
f_\alpha&P(\Sigma_1)^{\scriptscriptstyle T_1}\!\rightarrow\!
P(\Sigma_2)^{\scriptscriptstyle T_2}\!:\!A_1\!\mapsto\!T_2f(A_1\!\setminus\!K_1)
\,,\cr
g_\alpha&P(\Sigma_2)^{\scriptscriptstyle
T_2}\!\rightarrow\!P(\Sigma_2)^{\scriptscriptstyle
T_1}\!:\!A_2\!\mapsto\!K_1\!\cup\!f^{-1}(A_2)\,,\cr
\alpha_f&\Sigma_1\setminus\{\,p_1\in\Sigma_1\,|
\,f(p_1)=0_2\,\}\rightarrow\Sigma_2:p_1
\mapsto f(p_1)\,.\cr}}}}\par\medskip\par 
\noindent
Further, let  
$i:L\rightarrow P(\Sigma_{\scriptscriptstyle
L}):a\mapsto\{\,p\in\Sigma\,|\,p<a\,\}$ and 
$\pi:P(\Sigma_{\scriptscriptstyle L})\rightarrow L:A\mapsto\bigvee A\,$, 
with
${\cal L}:(\Sigma,T)\!\mapsto\!P(\Sigma)^{\scriptscriptstyle
T};\,\alpha\!\mapsto\!f_\alpha$ and
${\cal C}:L\!\mapsto\!(\Sigma_{\scriptscriptstyle
L},i\circ\pi);\,f\!\mapsto\!\alpha_f\,$. We obtain$\,$:
\begprop
\result $\pi\dashv i$ with associated simple closure 
$i\circ\pi:A\mapsto\{\,p\in\Sigma\,|\,p<\bigvee A\,\}\,$;
\result $\alpha_f$ is a morphism of closure spaces;
\result $f_\alpha\dashv g_\alpha$ is an adjunction of complete atomistic
lattices;
\result ${\cal L}\dashv{\cal C}$ is an equivalence between
\underbar{CSpace} and \underbar{JCALatt}$\,$.
\endprop 
\par\noindent Note that these results restrict in a canonical manner to
complete atomistic ortholattices. Indeed, for each such lattice let us
define the binary relation $\bot$ on $\Sigma_{\scriptscriptstyle L}$ by
$p\,\bot\,q$ if 
$p<q'\,$. Then the induced map
$T:P(\Sigma_{\scriptscriptstyle L})\rightarrow
P(\Sigma_{\scriptscriptstyle L}):A\mapsto A^{\bot\bot}$ with
$A^\bot=\{\,q\in\Sigma\,|\,(\forall p\in A)\,p\bot q\,\}$ is a simple
closure operator, with 
$A\subseteq\Sigma_{\scriptscriptstyle L}$ biorthogonal if and only if
$A=\{\,p\in\Sigma_{\scriptscriptstyle L}\,|\,p<\bigvee A\,\}\,$. Now  
$\bot$ is symmetric, $p\,\bot q\Rightarrow q\,\bot p\,$, 
antireflexive, $p\,\bot q\Rightarrow p\not=q\,$, and separating, 
$p\not=q\Rightarrow(\exists r\in\Sigma_{\scriptscriptstyle L})\,p\,\bot r\
\&\ q\,\bot\!\!\!\!/\,r\,$. In fact the last
condition is exactly the requirement  that singletons be biorthogonal. We
then obtain an equivalence between the  category \underbar{JCAoLatt} of
atomistic join complete ortholattices and  the category \underbar{OSpace}
of orthogonality spaces.

\minipar

The above considerations allow us to recover some standard results 
concerning power functors and complete Boolean algebras, results which 
will be generalised in the next section to transition structures. First,
let \underbar{CaBAlg} be the category of complete atomic Boolean algebras
with as morphisms all maps preserving the join, the meet, and the 
orthocomplementation. Note that any two of these conditions implies  the
third. Indeed, $\bigwedge A=(\,\bigvee\! A')'$ and 
$\bigvee A=(\,\bigwedge\! A')'$
so that any map which preserves the join and the orthocomplementation also
preserves the meet, whereas any map which preserves the meet and the 
orthocomplementation also preserves the join. On the other hand, if
$f$ preserves finite joins and meets and satisfies $f(0_1)=0_2$ and 
$f(1_1)=1_2\,$, then $f(a_1)\wedge f(a_1')=f(a_1\wedge a_1')=f(0_1)=0_2$ 
and $f(a_1)\vee f(a_1')=f(a_1\vee a_1')=f(1_1)=1_2\,$, so that 
$f(a_1')=f(a_1)'$ since any Boolean algebra is uniquely complemented.
Second, given the set function $\alpha:\Sigma_1\rightarrow\Sigma_2$ define 
$S_\alpha:P(\Sigma_2)\!\rightarrow\!P(\Sigma_1)
:A_2\!\mapsto\!\{x_1\in\Sigma_1\,|\,\alpha(x_1)\in A_2\}$ 
and
$P_\alpha:P(\Sigma_1)\!\rightarrow\!P(\Sigma_2)
:A_1\!\mapsto\!\{x_2\in\Sigma_2\,|\,(\exists x_1\in A_1)\,\alpha(x_1)=x_2\}\,$. 
Third, for $B$ a complete atomistic Boolean algebra let
$\mu_{\scriptscriptstyle B}:B\rightarrow P(\Sigma_{\scriptscriptstyle
B}):a\mapsto\{\,p\in\Sigma_{\scriptscriptstyle B}\,|\,p<a\,\}$  and
$\rho_{\scriptscriptstyle B}:P(\Sigma_{\scriptscriptstyle B})\rightarrow
B:A\mapsto\bigvee A\,$. Then for $f\dashv g$ an adjunction between
complete atomistic Boolean  algebras we obtain$\,$:
\begprop
\result $P_\alpha\dashv S_\alpha\,$, with $\alpha$ injective iff
$S_\alpha\circ P_\alpha={\rm id}$ or surjective iff $P_\alpha\circ
S_\alpha={\rm id}\,$;
\result $\mu_{\scriptscriptstyle B}\dashv\rho_{\scriptscriptstyle B}$ is
an equivalence, and $f(\Sigma_{\scriptscriptstyle
B_1})\subseteq\Sigma_{\scriptscriptstyle B_2}$ iff $g(a_2')=g(a_2)'$.
\endprop
\par\noindent 
The first result implies that the maps
$S:\hbox{\underbar{Set}}\rightarrow\hbox{\underbar{CaBalg}}^{\rm op}
:{\Sigma\mapsto P(\Sigma)}\,;\ \alpha\mapsto S_\alpha$  
and
$P:\hbox{\underbar{Set}}\rightarrow\hbox{\underbar{JCLatt}}
:\Sigma\mapsto P(\Sigma)\,;\ \alpha\mapsto P_\alpha$
are functors.
Indeed, $S$ is well defined,
$x_1\!\in\!S_\alpha(A_2^c)
\Leftrightarrow\alpha(x_1)\!\in\!A_2^c\Leftrightarrow\alpha(x_1)\!\notin\!A_2
\Leftrightarrow x_1\!\notin\!S_\alpha(A_2)\Leftrightarrow x_1\!\in\!
S_\alpha(A_2)^c$, and functorial, since we have
$x\in S_{{\rm id}}(A)\Leftrightarrow(\exists y\in A)\,x={\rm
id}(x)=y\Leftrightarrow x\in A$ and
$x_1\!\in\!S_{\alpha_2\circ\alpha_1}(A_3)\Leftrightarrow
(\alpha_2\!\circ\!\alpha_1)(x_1)\!\in\!
A_3\Leftrightarrow\alpha_1(x_1)\!\in\!S_{\alpha_2}(A_3)\Leftrightarrow
x_1\!\in\!(S_{\alpha_1}\!\circ\!S_{\alpha_2})(A_3)\,$. 
The second result implies that $S$ is an equivalence$\,$: by the first 
part the equivalence between closure spaces and complete atomistic 
lattices with join
preserving atomic maps restricts to an equivalence  between sets and
complete atomistic Boolean algebras with join preserving kernel free
atomic maps, whereas by the second part this latter category  is dual to
the category of complete atomistic Boolean algebras with maps  preserving
the meet, the orthocomplement, and so the join. 
\par\vskip 0.406 truecm\par
\par\vskip 0.406 truecm\par
\noindent   {\bf 6. Transition structures}    
\numprop=0\setcounter{section}{6}   
\par\vskip 0.406 truecm\par
\par\noindent    Now, in applying the power construction to complete
lattices we have two  relevant orders, namely $A\subseteq B$ in $P(L)$ and
$a<b$ in $L\,$. It is then of interest to consider the strong preorder on
$P(L)$ defined by $A\ll B$ if $\bigvee A<\bigvee B\,$. Note that this 
preorder is indeed superordinate to both of the original orders, since we have
that
$A\!\subseteq\!B\ \Rightarrow\
{\textstyle\bigvee}A\!<\!{\textstyle\bigvee}B\ \Rightarrow\ A\!\ll\!B$ and
$a\!<\!b\ \Rightarrow\
{\textstyle\bigvee}\{a\}\!=\!a\!<\!b\!=\!{\textstyle\bigvee}\{b\}\
\Rightarrow\ \{a\}\!\ll\!\{b\}\,$. Further, in the context of join
preserving maps the minimal element may be treated as redundant, since 
all such maps satisfy the condition $f(0_1)=0_2\,$. We are then lead to 
consider the truncated power set
$P_0(L)=P(\,L\setminus\{0\}\,)\,$. This reduction is reasonable, since
$P(L)=P(\,(L\setminus\{0\})\cup\{0\}\,)
=P(\,L\setminus\{0\}\,)\times P(\{0\})$ 
and 
$\bigvee(A\cup\{0\})=(\bigvee A)\vee(\bigvee\{0\})=(\bigvee
A)\vee0=\bigvee A\,$. In particular, we obtain the category
\underbar{PStruct} of power  structures,
$P\!S(L_1,L_2)=P_0(\,J(L_1,L_2)\,)\,$, and the category 
\underbar{FStruct} of functional structures, 
$F\!S(L_1,L_2)=J(\,P_0(L_1),P_0(L_2)\,)\,$.  Now in the category of power
structures we focus on the set of join  preserving maps $f:L_1\rightarrow
L_2\,$, whereas in the category of  functional structures we focus on the
set of union preserving  maps $\theta:P_0(L_1)\rightarrow P_0(L_2)\,$.
Since the condition of join preservation may be written $f(\,\bigvee
A_1)=\bigvee P_f(A_1)\,$, it is  then of interest to consider intermediate
structures defined by pairs 
$(f,\theta)$ satisfying the coherence condition 
$f\succ\theta\ \Leftrightarrow\ f(\bigvee A_1)=\bigvee\theta(A_1)$,
i.e., for the operational resolution 
$J:P_0(L)\rightarrow L:A\mapsto{\textstyle\bigvee}A$ we 
have $f\succ\theta\ \Leftrightarrow\ f\circ J_1=J_2\circ\theta\,$. Then,
for
$\ell:L\rightarrow P_0(L):a\mapsto(0,a]$:
\begprop
\result $J\dashv\ell$ with $J\circ\ell={\rm id}\,$;
\result Given union preserving $\theta\,$, if $f\succ\theta$ then $f$ is
unique and join preserving$\,$; 
\result If $\theta\,$ preserves unions, there exists 
$f\succ\theta$ iff $\theta$ is strongly isotone$\,$;
\result If $f_1\!\succ\!\theta_1$, $f_2\!\succ\!\theta_2$ then 
$f_2\!\circ\!f_1\!\succ\!\theta_2\!\circ\!\theta_1\,$; if 
$f_\alpha\!\succ\!\theta_\alpha$ then
$\vee_\alpha f_\alpha\!\succ\!\cup_\alpha\theta_\alpha\,$.
\endprop
\par\noindent 
The first result is a trivial
technical lemma allowing the compact  presentation of the coherence
condition. The second and third results allow the reconstruction of the
underlying morphism $f$ associated to a given covering $\theta\,$. The
first part of the fourth result allows us  to introduce the category
\underbar{TStruct} of transition structures,  where $T\!S(L_2,L_2)$ is the
set of pairs $(f,\theta)$ such that 
$f\succ\theta\,$, whereas the second part allows us to define the 
subcategory \underbar{BStruct} of based structures, whose morphisms are 
obtained by closing the Hom-sets of \underbar{PStruct} with respect to 
unions. We then obtain an inclusion hierarchy of intermediate categories
between \underbar{PStruct}$\,$, construed as an isomorphic image of
\underbar{JCLatt}$\,$, and \underbar{FStruct}$\,$, construed as the
maximal power construction on lattices.

\minipar

Note that the obvious inclusions
$\hbox{\underbar{PStruct}}\hookrightarrow\hbox{\underbar{BStruct}}
\hookrightarrow\hbox{\underbar{TStruct}}
\hookrightarrow\hbox{\underbar{FStruct}}$
are functorial, since the objects and compositions laws are the same in
the four categories. Further, the quantaloid morphism
$F:\hbox{\underbar{TStruct}}\rightarrow\hbox{\underbar{JCLatt}}
:(f,\theta)\mapsto f$ is retractive, since $F\circ{\rm I}={\rm id}$ for 
${\rm I}:{\underbar{JCLatt}}\rightarrow\hbox{\underbar{TStruct}}
:f\mapsto(f,P_f)\,$. Finally, the above
quantaloid inclusions are all strict, the  category of transition
structures then being the maximal faithful coherent enrichment of
\underbar{JCLatt}$\,$. Explicitly, let ${\bf2}$ be the two  element
lattice so that $P_0({\bf2})\!=\!P({\bf2}\setminus\{0\})\!=\!P(\{1\})\,$.
We then obtain$\,$:
\begprop
\result $P\!S({\bf2},L)=L$ whereas
$B\!S({\bf2},L)=T\!S({\bf2},L)=F\!S({\bf2},L)=P_0(L)\,$;
\result $P\!S(L,{\bf2})=B\!S(L,{\bf2})=T\!S(L,{\bf2})=L$ whereas
$F\!S(L,{\bf2})=P_0(L)\,$;
\result The inclusions
$\hbox{\underbar{PStruct}}\hookrightarrow\hbox{\underbar{BStruct}}
\hookrightarrow\hbox{\underbar{TStruct}}\hookrightarrow
\hbox{\underbar{FStruct}}\,$
are strict.
\endprop
\par
\noindent These results can be directly extended to partial constructs,
namely  categories ${\cal A}$ which are concrete over the category
\underbar{PSet} of sets with partially defined maps. Indeed, let us
consider a partial construct $U:{\cal A}\rightarrow\hbox{\underbar{PSet}}$
such that each  Hom-set contains a non-trivial morphism, ${\rm
Ker}(Uf)\not={\rm Dom}(Uf)\,$. First, the image of ${\cal P}\circ U\,$,
where
${\cal P}:\hbox{\underbar{PSet}}\rightarrow\hbox{\underbar{JCLatt}}$ is
the partial power functor, defines a category ${\cal P}{\cal A}$ which
generalises the category \underbar{PStruct} of power structures. Second,
by analogy to the category \underbar{BStruct} of based structures, closing
${\cal P}{\cal A}$ Hom-sets under pointwise unions we obtain the 
quantaloid ${\cal Q}^-{\cal A}\,$. Third, by analogy to the category 
\underbar{FStruct} of functional structures, taking all union preserving
maps we obtain the quantaloid ${\cal Q}^+{\cal A}\,$. Note that functors
$F:{\cal A}\rightarrow{\cal B}$ canonically lift to ${\cal P}$ and 
${\cal Q}^-$ but not to ${\cal Q}^+$. In a certain sense, then, we may
consider ${\cal Q}^-{\cal A}$ as a functorial enrichment of ${\cal A}$ and
${\cal Q}^+{\cal A}$ as its contextual enrichment. Now,  the above
inclusion hierarchy arises for the case 
$U:\hbox{\underbar{JCLatt}}\rightarrow\hbox{\underbar{PSet}}:L\mapsto
L\setminus\{0\}\,$. In particular, for any subquantaloid ${\cal A}$ of
\underbar{JCLatt} we may generalise the category \underbar{TStruct} of
transition structures to the quantaloid ${\cal Q}^0{\cal A}\,$, the map
$f$ being the property transition associated to the state transition
$\theta\,$. In fact, given any subcategory ${\cal A}$ of
\underbar{JCLatt}$\,$,  considering ${\cal P}{\cal A}$ as a subcategory of
\underbar{TStruct} we can define ${\cal E}{\cal A}\,$, the smallest
subquantale of
\underbar{JCLatt} containing ${\cal A}\,$, as the image of ${\cal P}{\cal
A}$ under the underlying functor. We then obtain pre-enrichments as free
extensions of subcategories guaranteeing that all ${\cal Q}^-$-morphisms
may be considered as state transitions.
\par\vskip 0.406 truecm\par
\par\vskip 0.406 truecm\par
\vfill
\eject
\noindent   {\bf 7. States and properties}      
\numprop=0\setcounter{section}{7}   
\par\vskip 0.406 truecm\par
\par\noindent    In this section we briefly indicate how the above
categorical techniques find a direct application in the so-called `Geneva
School approach', a framework theory allowing abstract mathematical
representations of concrete physical notions. The primitive concrete
notions of this approach are those  of `particular physical system',
namely a part of the ostensively external  phenomenal world considered as
distinct from its surroundings, and  `definite experimental project',
namely a real experimental procedure  where we have decided in advance
what would be the positive result should  we perform the experiment. We
then obtain the mathematical notions of `state', construed as an abstract
name for a singular realisation of the physical system, and `property',
construed as the element of reality  corresponding to a definite
experimental project. Now the sets $\Sigma$ of states and $L$ of properties
each possess mathematical structure arising directly from their
physical natures. First, two states
${\cal E}$ and ${\cal E}'$ are called orthogonal, written 
${\cal E}\bot\,{\cal E}'$, if there exists a definite experimental project
which is certain for the first and impossible for the second. The
orthogonality relation is then trivially symmetric and antireflexive. In
particular, writing $A^\bot$ for the orthogonal of $A\subseteq\Sigma$ we
have that $A\subseteq A^{\bot\bot}$ and $A\subseteq B\Rightarrow
B^\bot\subseteq A^\bot$. The map $A\mapsto A^{\bot\bot}$ is then a closure
operator, so that the set $(\Sigma,\bot)$ of biorthogonals is a complete
atomistic ortholattice with atoms $\{{\cal E}\}^{\bot\bot}$. Second, the
property $a$ is called stronger than the property $b\,$, written $a<b\,$,
if $b$ is actual  whenever $a$ is actual. The strength relation is then
trivially a partial order. Further, the construction of the product
$\Pi{\bf A}$ of the family
${\bf A}$ of definite experimental projects implies that $L$ is a 
complete lattice whose meet is given by semantic conjunction.  These two
structures are intimately linked by the so-called Cartan maps, which
associate to each property $a$ the set $\mu(a)$ of states in which it is
actual and to each state ${\cal E}$ the set $S({\cal E})$ of its actual
properties. Indeed, by definition
$\mu:L\rightarrow P(\Sigma)$ is injective and satisfies 
$\mu(\,\bigwedge\!A)=\bigcap\mu(A)\,$, whereas the induced map
$\rho:\Sigma\rightarrow L:{\cal E}\mapsto\bigwedge S({\cal E})$ satisfies
$a\!<\!b\Leftrightarrow
[\rho({\cal E})\!<\!a\Rightarrow\rho({\cal E})\!<\!b]\,$.
\par\vglue3pt

The standard axioms then allow one to characterise the images of $\mu$
and
$\rho\,$, namely those sets of states which represent properties and 
those properties which represent states.  First, note that each atom $p$
is a state representative. Indeed, $p\not=0$  so that there exists a state
${\cal E}$ in which $p$ is actual. Then 
$\rho({\cal E})<p$ so that $p=\rho({\cal E})\,$. Further, operationally
the  state represents all that can be done with certainty with the system.
The  general principle that the generation of any actual property requires
the  destruction of another then indicates that we should postulate a
bijective  correspondence between state representatives and atoms. The
property  lattice is then atomistic. Second, the orthogonality of two
states is an objective feature of the  system as a whole. We are then led
to suppose the existence of a map 
$^\sharp:\Sigma\rightarrow L:p\mapsto p^\sharp$ such that $p\,\bot\,q$ if
and only if $q<p^\sharp$, with induced action
$':L\rightarrow L:a\mapsto\bigwedge\{\,p^\sharp\,|\,p<a\,\}\,$. Then
$\mu(a')=\mu(a)^\bot$, and ${}'$ satisfies the conditions
$a<b\Rightarrow b'<a'$ and $a<a''$. Third, we can identify the lattices
$(\Sigma,\bot)$ and $L$ by supposing  in addition that $'$ be surjective,
that is, that each property have an opposite. Then the map ${}'$ is an
orthocomplementation, and if 
$p\not=q$ there exists $r$ such that $p\,\bot\,r$ and
$q\,\bot\!\!\!\!/\,\,r$. In summary, with these three axioms we can model
the set of properties  of any given physical system by a complete
atomistic ortholattice,  and the set of its states by an orthogonality
space, the physical state-property duality being a concrete realisation
of the abstract  equivalence between the categories \underbar{OSpace} and
\underbar{JCAoLatt}$\,$. Note that any complete atomistic ortholattice can
be putatively interpreted as a property lattice. First, in order to
interpret atoms as states, we assume that for each $p\in\Sigma$ there
exists a definite experimental project $\alpha_p$ which is certain for $q$
if and only if
$q=p\,$. Second, in order to interpret $q<p'$ as an orthogonality relation,
we assume that for each $p\in\Sigma$ there exists a definite experimental
project $\beta_p$ which is certain for $q$ if and only if $q<p'$. To
obtain
a coherent interpretation we then assume that $\beta_p=\alpha_p^\sim$, the
inverse obtained by exchanging the terms of the alternative. Completing
with respect to the product we then obtain exactly the lattice $L\,$,
since the condition $a=\bigwedge\{\,p'\,|\,p<a\,\}$ implies that $a'$ is
generated by the $\beta_p$ with $p$ majorised by $a\,$.
For a somewhat different categorical implementation of the Geneva School
axioms see [Valckenborgh 2000].
\par\vglue2pt

Apart from its mathematical elegance, the categorical realisation of 
state-property duality is a powerful tool in the study of axiomatic 
quantum theory. For example, it is implicit in the construction of many 
standard secondary notions, such as the classical decomposition, construed 
as a product in \underbar{JCAoLatt}$\,$, and observables, construed as
morphisms
in \underbar{COLatt} with domain a complete Boolean algebra whose Stone
space encodes the measurement scale. First, $z\in L$ is called
central if there exists a direct product decomposition 
$\pi:L_1\times L_2\simeq L$ with $z=\pi(1,0)\,$, in which case
$z$ has unique complement $z'=\pi(0,1)\,$. Now for complete atomistic
ortholattices, the central elements are exactly the classical properties,
namely those $z\in L$ such that for each $p\in\Sigma_{\scriptscriptstyle
L}$ either $p<z$ or $p<z'$. In particular, for $\alpha\in\Omega$ the atoms
of  the center of $L$ we have that $L=\times_\alpha[0,\alpha]\,$. Second,
the spectrum of the observable $\mu:{\bf B}\rightarrow L$ may
be obtained from the decomposition $N=\mu^\ast(0)\,$,
$D=\bigvee\{\,E\in\Sigma_{\scriptscriptstyle{\bf B}}\,|\,E<N'\,\}\,$,
$C=D'\wedge N'\,$, the interval $[0,D]$ being atomistic and so
representing
the discrete spectrum and the interval $[0,C]$ being atomless and so 
representing the continuous spectrum. Further, different physical and
mathematical aspects of a given problem are  often best treated using
different techniques, so that the various  categorical equivalences
exposed above are crucial in enabling natural  translations. For example,
the meet has a definite physical interpretation  as semantic conjunction,
being operationally constructible via the product,  whereas it is often
technically useful to consider join preserving maps,  these being the
usual objects for representation theorems. The duality  between
\underbar{MCLatt} and \underbar{JCLatt} then allows one to define  a
concept in physically meaningful terms and then study it using 
mathematically adequate techniques. For example, the construction of Hilbertian
realisations for Arguesian orthogeometries $(G,\oplus)$ is  purely
categorical, being based on an appropriate embedding of $G$ as a
hyperplane in the projective geometry $\overline{G}$ whose elements are 
endomorphisms of $G$ with a given fixed axis $H\,$. First, the set
$V=\overline{G}\setminus G$ has a natural vector space structure over the 
division ring of homotheties defined with respect to a fixed $0\in V\,$. 
Second, each non-degenerate morphism $g:G_1\setminus N_1\rightarrow G_2$ 
admits a canonical extension 
$h:\overline{G}{}_1\setminus\overline{N}{}_1\rightarrow\overline{G}_2$
whose restriction to $V_1$ defines a semilinear map $f:V_1\rightarrow
V_2\,$. Third, and finally, the orthogonality relation defines a 
non-degenerate homomorphism
$g:G\rightarrow G^\ast:p\mapsto\{p\}^\bot$, where $G^\ast$ is the dual
geometry of hyperplanes of $G\,$, whose associated quasilinear map
$f:V\rightarrow V^\ast$ then defines a Hermitian form.
\par\vglue3pt  

A case in point is the notion of deterministic evolutions. First, one can
partially encode an imposed evolution by defining a map $\Phi_{01}$ which 
associates to each definite experimental project $\alpha_1$ defined at 
the final time $t_1$ the definite experimental project $\alpha_0$ defined 
at the initial time $t_0$ by the prescription `Evolve the system as
required  and effectuate $\alpha_1\,$' [Daniel 1982, 1989]. Then it can be
physically  demonstrated that $\Phi_{01}$ preserves the product and maps $O_1$
to $O_0\,$.  We thereby obtain an induced map $\phi_{01}:L_1\rightarrow
L_0$ which preserves non-empty meets and maps $0_1$ to $0_0\,$. In words,
each given evolution  induces a weak balanced meet morphism. In the
simplest case, where
$\phi_{01}(1_1)=1_0$ so that the system is never destroyed by the
evolution, we then obtain the dense right adjoint
$\psi_{10}:L_0\rightarrow L_1\,$. Physically, the adjunction condition
implies that $\psi_{10}(p_0)$ is the strongest final property whose
actuality is guaranteed by the evolution for the 
initial state defined by  
the atom $p_0\,$. If this property in fact determines the final state of the
system, then we can realise the evolution  as a dense atomic join
morphism. Finally, under suitable stability conditions two orthogonal
final states arise from two orthogonal initial states, since if $\alpha_1$
separates $\psi_{10}(p_0)$ and $\psi_{10}(q_0)$ then
$\Phi_{01}\alpha_1$ separates $p_0$ and $q_0\,$. In the Hilbertian context, 
one can then prove that a sufficiently continuous evolution may be
represented by a unitary flow [Faure, Moore and Piron 1995]. More generally,
the cognitive duality between causal assignment and consecutive
propagation  of properties may be encoded in the quantaloid isomorphism
$A^\ast:\hbox{\underbar{JCLatt}}\rightarrow\hbox{\underbar{MCLatt}}^{\rm coop}$
[Coecke 2000; Coecke, Moore and Stubbe 2000]. Explicitly, given two 
property lattices $L_1$ and $L_2$ let us write $a_1\to\hspace{-2mm}\to a_2$ if
$a_1\in L_1$ is a material cause of $a_2\in L_2\,$. Then by the
operational signification of the lattice partial order as semantic
implication and the operational construction of the lattice meet as
semantic conjunction, the relation
$\to\hspace{-2mm}\to\,$ should be fully isotone,    
$(x_1<a_1\to\hspace{-2mm}\to a_2<x_2)\Rightarrow(x_1\to\hspace{-2mm}\to x_2)\,$, and preserve
right non-empty meets,
$(a_1\to\hspace{-2mm}\to a_{2\alpha})\Rightarrow(a_1\to\hspace{-2mm}\to\wedge_\alpha
a_{2\alpha})\,$.  
Setting $g:L_2\rightarrow L_1:a_2\mapsto\bigvee\{\,a_1\in
L_1\,|\,a_1\to\hspace{-2mm}\to a_2\,\}$ we may then express causal assignment by a
weak meet morphism, in the sense that $(a_1\to\hspace{-2mm}\to
a_2)\Leftrightarrow(a_1<g(a_2))\,$,  whose pseudoadjoint
$f:L_1^u\rightarrow L_2^u$ implements the  notion of
consecutive propagation of properties. In particular, the  interpretation
of propagation as an evolutive flow may be complemented by the
interpretation of assignment in terms of states of compoundness.
\par\vskip 0.406 truecm\par
\par\vskip 0.406 truecm\par

\vfill
\eject
\noindent{\bf APPENDIX}    
\numprop=0   
\par\vskip 0.406 truecm\par\noindent 
In this appendix we provide proofs of the results cited in the text.

\bigskip\noindent   
\prf{3.1}
\par\noindent (1) Let ${\rm id}_1<(g\circ f)$ and $(f\circ g)<{\rm
id}_2\,$. Then
$f(a_1)<a_2\Rightarrow a_1<(g\circ f)(a_1)<g(a_2)$ and
$a_1<g(a_2)\Rightarrow f(a_1)<(f\circ g)(a_2)<a_2\,$. Let
$f(a_1)<a_2\Leftrightarrow a_1<g(a_2)\,$.  Then for each $a_1\in L_1$ we
have that 
$f(a_1)<f(a_1)\Rightarrow a_1<(g\circ f)(a_1)$ and for each $a_2\in L_2$
we have that
$g(a_2)<g(a_2)\Rightarrow(f\circ g)(a_2)<a_2\,$.
\par\noindent 
(2) First $f^{\!-1}\!=\!g\!\Leftrightarrow\!g\!\circ\!f\!=\!{\rm id}_1\,$; 
second $f\!\circ\!g\!=\!{\rm id}_2\!\Leftrightarrow\!{\rm
id}_1\!<\!g\!\circ\!f,f\!\circ\!g\!<\!{\rm id}_2\,$;
third ${\rm
id}_2\!<\!f\!\circ\!g,g\!\circ\!f\!<\!{\rm
id}_1\!\Leftrightarrow\!f\!\dashv\!g\!\dashv\!f$. 
\par\noindent (3) ${\rm id}_1<g\circ f$ so that $f=f\circ{\rm id}_1<f\circ
g\circ f$ and $g={\rm id}_1\circ g<g\circ f\circ g\,$, whereas
$f\circ g<{\rm id}_2$ so that $f\circ g\circ f<{\rm id}_2\circ f=f$ and
$g\circ f\circ g<g\circ{\rm id}_2=g\,$.
\par\noindent (4) Let $f<\overline{f}\,$. Then 
$(f\circ\overline{g}\,)(a_2)<(\,\overline{f}\circ\overline{g}\,)
(a_2)<a_2\Rightarrow\overline{g}(a_2)<g(a_2)$
for each $a_2\in L_2\,$. Let $\overline{g}<g\,$. Then 
$a_1<(\,\overline{g}\circ\overline{f}\,)(a_1)<(g\circ\overline{f}\,)
(a_1)\Rightarrow
f(a_1)<\overline{f}(a_1)$ for each $a_1\in L_1\,$.
\par\noindent (5) ${\rm id}(a)<b\Leftrightarrow a<b\Leftrightarrow a<{\rm
id}(b)\,$, and so ${\rm id}\dashv{\rm id}\,$. Let $f\dashv g$ and
$\overline{f}\dashv\overline{g}\,$. Then
$(\,\overline{f}\circ f)(a_1)<a_3\Leftrightarrow
f(a_1)<\overline{g}(a_3)\Leftrightarrow
a_1<(g\circ\overline{g}\,)(a_3)\,$, and so $(\,\overline{f}\circ
f)\dashv(g\circ\overline{g}\,)\,$.    
\par\bigskip\par\noindent
\prf{3.2} 
\par\noindent (1)
$f(\bigvee\!A_1)\!<\!a_2\!\Leftrightarrow\!\bigvee\!A_1\!<\!g(a_2)
\!\Leftrightarrow\!(\forall
a_1\!\!\in\!\!A_1)\,[a_1\!\!<\!\!g(a_2)\!\!
\Leftrightarrow\!\!f(a_1)\!\!<\!\!a_2]
\!\Leftrightarrow\!\bigvee\!f(A_1)\!<\!a_2$;
$a_1\!<\!g(\bigwedge\!A_2)\!\Leftrightarrow\!f(a_1)\!<\!\bigwedge
\!A_2\!\Leftrightarrow\!(\forall
a_2\!\in\!A_2)\,[f(a_1)\!\!<\!\!a_2\!\!\Leftrightarrow\!\!a_1\!\!<\!\!g(a_2)]
\!\Leftrightarrow\!a_1\!<\!\bigwedge\!g(A_2)$.
\par\noindent (2) 
If $f(\bigvee\!A)=\bigvee\!f(A)$ then  
$f(a_1)\!<\!a_2\Rightarrow a_1\!<\!\bigvee\{\,x_1\!\in\!L_1\,|\,f(x_1)\!<\!a_2\}
\!=\!f^\ast(a_2)$ and
$a_1\!<\!f^\ast(a_2)\Rightarrow f(a_1)\!<\!f(\bigvee\{\,x_1\!\in\!
L_1\,|\,f(x_1)\!<\!a_2\,\})=\bigvee\{\,f(x_1)\,|\,f(x_1)\!<\!a_2\,\}\!<\!a_2$.
\par\noindent (3) 
If $g(\bigwedge A)=\bigwedge g(A)$ then 
$a_1\!<\!g(a_2)\Rightarrow g_\ast(a_1)\!=\!\bigwedge\{\,x_2\!\in\!
L_2\,|\,a_1\!<\!g(x_2)\,\}\!<\!a_2$ and
$g_\ast(a_1)\!<\!a_2\Rightarrow
a_1\!<\!\bigwedge\{\,g(x_2)\,|\,a_1\!<\!g(x_2)\,\}=g(\bigwedge\{\,x_2\!\in\!
L_2\,|\,a_1\!<\!g(x_2)\,\}\!<\!g(a_2)$.
\par\noindent (4) $(\vee_\alpha
f_\alpha)(a_1)<a_2\Leftrightarrow(\forall\alpha)\,f_\alpha(a_1)<a_2
\Leftrightarrow(\forall\alpha)\,a_1<g_\alpha(a_2)\Leftrightarrow
a_1<(\wedge_\alpha g_\alpha)(a_2)$.
\par\noindent (5) $(f\!\circ\!\vee_\alpha
f_\alpha)(a_1)\!<\!a\Leftrightarrow(\forall\alpha)\,[f_\alpha(a_1)\!<\!g(a)
\!\Leftrightarrow\!a_1\!<\!(g_\alpha\!\circ\!g)(a)]\Leftrightarrow
a_1<\wedge_\alpha(g_\alpha\!\circ\!g)(a)$; $(\vee_\alpha
f_\alpha\!\circ\!f)(a)\!<\!a_2\Leftrightarrow(\forall\alpha)
\,[f(a)\!<\!g_\alpha(a_2)\!\Leftrightarrow\!
a\!<\!(g\!\circ\!g_\alpha)(a_2)]\Leftrightarrow
a<\wedge_\alpha(g\!\circ\!g_\alpha)(a_2)$.
\par\bigskip\par\noindent
\prf{3.3}
\par\noindent (1) First, let
$A^\ast:J(L_1,L_2)\!\rightarrow\!M(L_2,L_1)^{\rm co}:f\!\rightarrow\!f^\ast$. 
Then $A^\ast$ is well defined, since $f^\ast$ is unique and 
preserves meets, and isotone, 
$f\!\!<\!\!\overline{f}\!\Rightarrow\!\overline{f}{\,}^\ast\!\!<\!\!f^\ast
\!\Rightarrow\!f^\ast\!\!<^{\rm op}\!\!\overline{f}{\,}^\ast$. Second, let
$A_\ast:M(L_2,L_1)^{\rm co}\rightarrow J(L_1,L_2):g\mapsto g_\ast\,$. 
Then $A_\ast$ is well defined, since $g_\ast$ is unique and 
preserves joins, and isotone, 
$g<^{\rm op}\overline{g}\ \Rightarrow\ \overline{g}<g\ \Rightarrow\
g_\ast<\overline{g}{}_\ast\,$. Third, $A_\ast\circ A^\ast={\rm id}$ and
$A^\ast\circ A_\ast={\rm id}\,$.  Indeed, if $f\dashv g$ then $g=f^\ast$
and $f=g_\ast\,$. Hence 
$(A_\ast\circ A^\ast)(f)=(f^\ast)_\ast=g_\ast=f$  and
$(A^\ast\circ A_\ast)(g)=(g_\ast)^\ast=f^\ast=g\,$.
\par\noindent (2) First, let
$A^\ast:\hbox{\underbar{JCLatt}}\rightarrow\hbox{\underbar{MCLatt}}{}^{\rm
op}:L\mapsto L\,;\ f\mapsto f^\ast$. Then $A^\ast$ is well defined, since
we have $f\in J(L_1,L_2)\ \Rightarrow\ f^\ast\in M(L_2,L_1)=M^{\rm
op}(L_1,L_2)\,$,
and a functor, 
$A^\ast(f_2\circ f_1)=(f_2\circ f_1)^\ast=f_1^\ast\circ
f_2^\ast=f_2^\ast\circ^{\rm op}f_1^\ast=A^\ast(f_2)\circ^{\rm
op}A^\ast(f_1)\,$. Second, let
$A_\ast:\hbox{\underbar{MCLatt}}{}^{\rm
op}\rightarrow\hbox{\underbar{JCLatt}}:L\mapsto L\,;\ g\mapsto g_\ast\,$.
Then $A_\ast$ is well defined, since we have
$g\in M^{\rm op}(L_1,L_2)=M(L_2,L_1)\ \Rightarrow\ g_\ast\in
J(L_2,L_2)\,$,
and a functor,
$A_\ast(g_2\circ^{\rm op}g_1)=A_\ast(g_1\circ g_2)=(g_1\circ
g_2)_\ast=g_{2\ast}\circ g_{1\ast}=A_\ast(g_2)\circ A_\ast(g_1)\,$. Third,
as above $A_\ast\circ A^\ast={\rm id}$ and $A^\ast\circ A_\ast={\rm
id}\,$.
\par\noindent (3) Let
$A^\ast\!:\!\hbox{\underbar{JCLatt}}\!\rightarrow\!
\hbox{\underbar{MCLatt}}{}^{\rm coop}\!:\!f\!\mapsto\!f^\ast$
and
$A_\ast\!:\!\hbox{\underbar{MCLatt}}{}^{\rm coop}\!\rightarrow
\!\hbox{\underbar{JCLatt}}\!:\,g\!\mapsto\!g_\ast\,$.
Then combining the above two results we have that $A^\ast$ and $A_\ast$
are well  defined, are quantaloid morphisms, and satisfy $A_\ast\circ
A^\ast={\rm id}$ and 
$A^\ast\circ A_\ast={\rm id}\,$.
\par\bigskip\par\noindent
\prf{3.4}
\par\noindent (1)
$a_1\!<\!b_1\Rightarrow b_1'\!<\!a_1'
\Rightarrow\alpha(b_1')\!<\!\alpha(a_1')\Rightarrow
(C\alpha)(a_1)\!=\!\alpha(a_1')'\!<\!\alpha(b_1')'\!=\!(C\alpha)(b_1)\,$,
so that $C(\alpha)$ is isotone and $C$ is well defined.
$(C\,{\rm id})(a)={\rm id}(a')'=a''=a$ and
$C(\alpha_2\circ\alpha_1)(a_1)\!=\!(\alpha_2\circ\alpha_1)(a_1')'\!=
\!\alpha_2(\alpha_1(a_1')'')'\alpha_2((C\alpha_1)(a_1)')'\!=\!(C\alpha_2\circ
C\alpha_1)(a_1)$, so that $C$ is a functor.
$(CC\alpha)(a_1)\!=\!(C\alpha)(a_1')'\!=\!\alpha(a_1'')''\!=\!\alpha(a_1)$.
If $f\!\dashv\!g$ then
$(Cg)(a_2)\!=\!g(a_2')'\!<\!a_1\!\Leftrightarrow\!a_1'\!<\!g(a_2')\!
\Leftrightarrow\!f(a_1')\!<\!a_2'\!\Leftrightarrow\!a_2\!<\!f(a_1')'
\!=\!(Cf)(a_1)\,$,
so that $C(g)\dashv C(f)\,$. Hence $C$ maps join maps to meet 
maps and conversely. 
\par\noindent (2) 
First, if $f\dashv g$ then $f^\dagger=C(g)\dashv
C(f)=g_\dagger$ and so $\dagger$  is well defined on
\underbar{JCoLatt}$\,$. Second, 
$(f_2\circ f_1)^\dagger=C((f_2\circ f_1)^\ast)=C(f_2{}^\ast\circ
f_1{}^\ast)=C(f_1{}^\ast)\circ C(f_2{}^\ast)=f_1{}^\dagger\circ
f_2{}^\dagger$ and
$f^{\dagger\dagger}=C(f^{\dagger\ast})=C(C(f)_\ast{}^\ast)=(CC)(f)=f\,$,
so that $\dagger$ is an involution. Third, by the adjunction condition
$f^\dagger(a_2)\!<\!a_1'\Leftrightarrow
a_1\!<\!f^\dagger(a_2)'\!=\!f^\ast(a_2')
\Leftrightarrow f(a_1)\!<\!a_2'\,$. Fourth,
$f^\dagger\!\circ\!f\!=\!0_1\!\Leftrightarrow\!(f^\dagger\!\circ\!
f)(1_1)\!<\!0_1\!\Leftrightarrow\!f(1_1)\!=\!f(0_1')\!<\!f(1_1)'
\!\Leftrightarrow\!f(1_1)\!=\!0_2\!\Leftrightarrow\!f\!=\!0_2\,$.
\par\noindent (3) 
If $u^\dagger\!\circ\!u\!=\!{\rm id}$ then
$a_1\!<\!b_1'\!\Leftrightarrow\!(u^\dagger\!\circ\!u)(a_1)\!=\!a_1\!<\!b_1'
\!\Leftrightarrow\!u(b_1)\!<\!u(a_1)'\!\Leftrightarrow\!u(a_1)\!<\!u(b_1)'$. 
If conversely then
$a_1\!<\!x_1\!\Leftrightarrow\!u(a_1)\!<\!u(x_1')'\!\Leftrightarrow
\!u(x_1')\!<\!u(a_1)'\!\Leftrightarrow\!(u^\dagger\circ
u)(a_1)\!<\!x_1$. 
\par\noindent (4) First,
$h_\ast(a_2')\!\!<\!\!a_1\!\Leftrightarrow\!a_2'\!\!<\!\!h(a_1)
\!\Leftrightarrow\!h(a_1')\!\!=\!\!h(a_1)'\!\!<\!\!a_2
\!\Leftrightarrow\!a_1'\!\!<\!\!h^\ast(a_2)
\!\Leftrightarrow\!h^\ast(a_2)'\!\!<\!\!a_1$.
Second, $h^\dagger(a_2)\!\!=\!\!h^\ast(a_2')'\!\!=\!\!h_\ast(a_2)''
\!\!=\!\!h_\ast(a_2)$ and
$h_\dagger(a_2)\!\!=\!\!h_\ast(a_2')'\!\!=
\!\!h^\ast(a_2)''\!\!=\!\!h^\ast(a_2)$. 
Third, $h\circ h^\dagger\circ h=h\circ h_\ast\circ h=h$ and 
$h\circ h_\dagger\circ h=h\circ h^\ast\circ h=h\,$.
\par\bigskip\par\noindent
\prf{4.1}
\par\noindent (1) First, if $\chi=0$ then 
$0<C^a(x)$ and $\alpha_a(0)=0<x$ whereas if $\chi=1$ then 
$1<C^a(x)\Leftrightarrow C^a(x)=1\Leftrightarrow\alpha_a(1)=a<x\,$. Hence
$\alpha_a\dashv C^a$. Second, we have
$(f\circ\alpha_{a_1})(0)=f(0_1)=0_2=\alpha_{f(a_1)}(0)$ and
$(f\circ\alpha_{a_1})(1)=f(a_1)=\alpha_{f(a_1)}(1)\,$. 
Third, we have
$(C^{a_1}\circ g)(a_2)=1\Leftrightarrow a_1<g(a_1)\Leftrightarrow
f(a_1)<a_2\Leftrightarrow C^{f(a_1)}(a_2)=1\,$. 
\par\noindent (2) First, if $\chi=0$ then 
$x<\alpha^a(0)=a\Leftrightarrow C_a(x)=0\Leftrightarrow C_a(x)<0$
whereas if
$\chi=1$ then  
$x<1=\alpha^a(x)$ and $C_a(x)<1\,$. Hence $C_a\dashv\alpha^a$.
Second, we have  $(g\circ\alpha^{a_2})(0)=g(a_2)=\alpha^{g(a_2)}(0)$ and
$(g\circ\alpha^{a_2})(1)=g(1_2)=1_1=\alpha^{g(a_2)}(1)\,$.
Third, we have
$(C_{a_2}\circ f)(a_1)=0\Leftrightarrow f(a_1)<a_2\Leftrightarrow
a_1<g(a_2)\Leftrightarrow C_{g(a_2)}(a_1)=0\,$.
\par\noindent (3) Let $x\in L\,$. Then 
$(i_a\circ\pi_a)(x)=i_a(x\wedge a)=x\wedge a<x$ and
$i_a\circ\pi_a<{\rm id}\,$. Let $x\in[0,a]\,$. Then
$x=x\wedge a=\pi_a(x)=(\pi_a\circ i_a)(x)$ and $\pi_a\circ i_a={\rm
id}\,$. Let $x\in L\,$. Then
$x<[\,x\,(x<a)\,:\ 1\,(x\not<a)\,]
=[\,\hat{\imath}{}_a(x)\,(x<a)\,;\
\hat{\imath}{}_a(a)\,(x\not<a)\,]=(\hat{\imath}{}_a\circ\hat{\pi}{}_a)(x)$
and ${\rm id}<\hat{\imath}{}_a\circ\hat{\pi}{}_a\,$. Let
$x\in[0,a]\,$. Then 
$(\hat{\pi}{}_a\circ\hat{\imath}{}_a)(x)=[\,\hat{\pi}{}_a(x)\,(x\not=a)\,;\
\hat{\pi}{}_a(1)\,(x=a)\,]=x$ and
$\hat{\pi}{}_a\circ\hat{\imath}{}_a={\rm id}\,$.
%\vfill
%\eject
%%%%%%%%%%
%%%%%%%%%%
%%%%%%%%%%
\par\bigskip\par\noindent
\prf{4.2}
\par\noindent (1) Let $f$ be balanced. Then 
$g(a_2)\!=\!1_1\Leftrightarrow1_1\!<\!g(a_2)\Leftrightarrow
1_2\!=\!f(1_1)\!<\!a_2\Leftrightarrow a_2\!=\!1$, and $g$ is dense. 
Let $g$ be dense. Then 
$1_1\!<\!(g\circ f)(1_1)\Rightarrow f(1_1)\!=\!1_2$, and $f$ is  balanced. Let
$f$ be dense. Then $(f\circ g)(0_2)\!<\!0_2\Rightarrow g(0_2)\!=\!0_1$, and 
$g$ is balanced.  Let $g$ be balanced. Then
$f(a_1)=0_2\Leftrightarrow f(a_1)\!<\!0_2\Leftrightarrow
a_1\!<\!g(0_2)=0_1\Leftrightarrow a_1=0_1$, and $f$ is dense.
\par\noindent (2) Let $f$ be epic. Then 
$C_{a_2}\circ f=C_{g(a_2)}=C_{(g\circ f\circ g)(a_2)}
=C_{(f\circ g)(a_2)}\circ f\,$, so that 
$C_{a_2}=C_{(f\circ g)(a_2)}$ and $a_2=(f\circ g)(a_2)\,$. 
Let $f$ be surjective. Then with $a_2=f(a_1)$ we have
$(f\circ g)(a_2)=(f\circ g\circ f)(a_1)=f(a_1)=a_2$ and 
$f\circ g={\rm id}_2\,$. Let $f\circ g={\rm id}_2\,$. Then 
$g(a_2)=g(b_2)\Rightarrow
a_2={\rm id}_2(a_2)=(f\circ g)(a_2)=(f\circ g)(b_2)={\rm id}_2(b_2)=b_2\,$.
Let $g$ be injective. Then
$g\circ g_1=g\circ g_2\Rightarrow
(g\circ g_1)(a)=(g\circ g_2)(a)\Rightarrow g_1(a)=g_2(a)\,$.
Let $g$ be monic. Then 
$f_1\circ f=f_2\circ f\Rightarrow
g\circ g_1=(f_1\circ f)^\ast=(f_2\circ f)^\ast=g\circ g_1\,$, so that
$g_1=g_2\Rightarrow f_1=g_{1\ast}=g_{2\ast}=f_2\,$. 
\par\noindent (3) Let $f$ be a monic. Then 
$f(a_1)\!=\!f(b_1)\Rightarrow
f\!\circ\!\alpha_{a_1}\!=\!\alpha_{f(a_1)}\!=\!\alpha_{f(b_1)}
\!=\!f\!\circ\!\alpha_{b_1}$,
so that $\alpha_{a_1}\!=\!\alpha_{b_1}$ and $a_1\!=\!b_1\,$. 
Let $f$ be injective. Then 
$(f\circ g\circ f)(a_1)=f(a_1)$ and $(g\circ f)(a_1)=a_1\,$. 
Let $g\!\circ\!f\!=\!{\rm id}_1$. Then 
$a_1\!=\!{\rm id}_1(a_1)\!=\!(g\!\circ\!f)(a_1)$. Let
$g$ be surjective. Then 
$g_1\circ g=g_2\circ g\Rightarrow 
g_1(a_1)=(g_1\circ g)(a_2)=(g_2\circ g)(a_2)=g_2(a_1)$
for $a_1=g(a_2)$. Let $g$ be epic. Then
$f\circ f_1=f\circ f_2\Rightarrow 
g_1\circ g=(f\circ f_1)^\ast=(f\circ f_2)^\ast=g_2\circ g\,$, so that
$g_1=g_2\Rightarrow f_1=g_{1\ast}=g_{2\ast}=f_2\,$. 
\par\bigskip\par\noindent
\prf{4.3}
\par\noindent (1) If $b\!<\!\Pi_\beta((a_\alpha))$ then
$\alpha\!=\!\beta\Rightarrow(i_\beta(b))_\alpha\!=\!b\!<\!a_\alpha$ and 
$\alpha\!\not=\!\beta\Rightarrow(i_\beta(b))_\alpha\!=\!0\!<\!a_\alpha$.  
If $i_\beta(b)<(a_\alpha)$ then 
$b=(i_\beta(b))_\beta<a_\beta=\Pi_\beta((a_\alpha))$. 
Hence $i_\beta\dashv\Pi_\beta\,$. If $(a_\alpha)<j_\beta(b)$ then 
$\Pi_\beta((a_\alpha))\!=\!a_\beta\!<\!(j_\beta(b))_\beta\!=\!b$. If
$\Pi_\beta((a_\alpha))\!<\!b$ then 
$\alpha\!=\!\beta\Rightarrow a_\alpha\!<\!b\!=\!(j_\beta(b))_\alpha$ 
and $\alpha\!\not=\!\beta\Rightarrow a_\alpha\!<\!1\!=\!(j_\beta(b))_\alpha$. 
Hence $\Pi_\beta\dashv j_\beta\,$.
If $x<{\rm I}_\beta(b)=b$ then $x\in L_\beta$ and 
$\sigma_\beta(x)=x<b\,$. If $\sigma_\beta<b$ then 
$x\in L_\beta\Rightarrow x=\sigma_\beta(x)<b={\rm I}_\beta(b)$ and 
$x\notin L_\beta\Rightarrow 1=\sigma_\beta(x)<b\Rightarrow 
x<1=b={\rm I}_\beta(b)\,$. Hence $\sigma_\beta\dashv{\rm I}_\beta\,$.
If $b<\rho_\beta(x)$ then 
$x\!\in\!L_\beta\Rightarrow{\rm I}_\beta(b)\!=\!b\!<\!\rho_\beta(x)\!=\!x$ 
and $x\!\notin\!L_\beta\Rightarrow b\!<\!\rho_\beta(x)\!=\!0
\Rightarrow{\rm I}_\beta(b)\!=\!b\!=\!0\!<\!x$. If $b={\rm I}_\beta(b)<x$ 
then $x\in L_\beta$ and $b<x=\rho_\beta(x)\,$. 
Hence ${\rm I}_\beta\dashv\rho_\beta\,$.
\par\noindent (2) Let $p_\beta:L\rightarrow L_\beta$ be maps. Now 
$a\in\times_\alpha L_\alpha\Rightarrow a=(\Pi_\alpha(a))\,$. 
Hence the unique map 
$\theta:L\rightarrow\times_\alpha L_\alpha$ with
$p_\beta=\Pi_\beta\circ\theta$ is 
$\theta(x)=((\Pi_\alpha\circ\theta)(x))=(p_\alpha(x))\,$. Since the
partial order, minimal and maximal elements, joins and meets   are
computed pointwise, $\theta$ will be a morphism in 
\underbar{BPos}$\,$, \underbar{JCLatt} or \underbar{MCLatt} iff
each $p_\alpha$ is. 
\par\noindent (3) First, let $q_\beta\!:\!L_\beta\!\rightarrow\!L$ be
morphisms. Then the  unique map $\phi\!:\!\dot{\cup}_\alpha
L_\alpha\!\rightarrow\!L$ such that 
$q_\beta=\phi\circ {\rm I}_\beta$ is given by
$\phi(b)=(\phi\circ {\rm I}_\beta)(b)=q_\beta(b)$ for $b\in L_\beta\,$.
Note that there is no ambiguity for $x=0$ or $x=1$ since $q_\beta(0)=0$ 
and $q_\beta(1)=1\,$. It remains to prove that $\phi$ is isotone. Let 
$x<x^\ast\,$. Then $x$ and $x^\ast$ belong to the same component, say 
$L_\beta\,$, and $\phi(x)=q_\beta(x)<q_\beta(x^\ast)=\phi(x^\ast)\,$.
Second, $\Pi_\beta$ is the product in \underbar{MCLatt}$\,$, so that
$\Pi_\beta$  is the coproduct in \underbar{MCLatt}$^{\rm op}$ and
$i_\beta$ is the  coproduct in \underbar{JCLatt}$\,$.  Third, $\Pi_\beta$
is the product in \underbar{JCLatt}$\,$, so that $\Pi_\beta$  is the
coproduct in \underbar{JCLatt}$^{\rm op}$ and $j_\beta$ is the  coproduct
in \underbar{MCLatt}$\,$. 
%\vfill
%\eject
%%%%%%%%%%
%%%%%%%%%%
%%%%%%%%%%
\par\bigskip\par\noindent 
\prf{4.4}
\par\noindent (1) First, for $x_2\not={\bf1}$ we have
$(F_\alpha\circ
G_\alpha)(x_2)=F_\alpha(\alpha^\ast(x_2))=
(\alpha\circ\alpha^\ast)(x_2)<x_2\,$,
and for $x_2={\bf 1}$ we have 
$(F_\alpha\circ G_\alpha)({\bf1})=F_\alpha({\bf1})={\bf1}\,$.  On the
other hand, for $x_1<a_1$ we have 
$x_1<(\alpha^\ast\circ\alpha)(x_1)=G_\alpha(\alpha(x_1))=(G_\alpha\circ
F_\alpha)(x_1)=G_\alpha\,$, and for $x_1\not<a_1$ we have 
$x_1<{\bf1}=G_\alpha({\bf1})=(G_\alpha\circ F_\alpha)\,$. Second, we have
that
$(F_{\alpha_2}\!\circ\!F_{\alpha_1})(x_1)\!=\!{\bf1}$  iff
$x_1\!\not<\!(G_{\alpha_1}\!\circ\!G_{\alpha_2})(1_3))\!=
\!G_{\alpha_1}(\alpha_2^\ast(1_3))\!=\!G_{\alpha_1}(a_2)\!=
\!\alpha_1^\ast(a_2)$
iff $F_{\alpha_2\circ\alpha_1}(x_1)={\bf1}\,$, and for 
$x_1<\alpha^\ast(a_2)$ we have
$F_{\alpha_2\circ\alpha_1}(x_1)=(\alpha_2\circ\alpha_1)(x_1)=
(F_{\alpha_2}\circ
F_{\alpha_2})(x_1)\,$. Third, for $x_1<F^\ast(1_2)$ we have 
$F_{\alpha_F}(x_1)=\alpha_{\scriptscriptstyle F}(x_1)=F(x_1)$, and for
$x_1\not<F^\ast(1_2)$ we have
$F_{\alpha_F}(x_1)={\bf1}=F(x_1)$ since
$x_1\not<F^\ast(1_2)\Leftrightarrow F(x_1)\not<1_2\Leftrightarrow
F(x_1)={\bf1}\,$.
\par\noindent (2) First, for $x_2\in L_2$ we have
$(\alpha_{\scriptscriptstyle F}\circ\beta_{\scriptscriptstyle
F})(x_2)=\alpha_{\scriptscriptstyle F}(F^\ast(x_2))=(F\circ
F^\ast)(x_2)<x_2$, and for $x_1<F^\ast(1_2)$ we have 
$x_1<(F^\ast\circ F)(x_1)=\beta_{\scriptscriptstyle
F}(F(x_1))=(\beta_{\scriptscriptstyle F}\circ\alpha_{\scriptscriptstyle
F})(x_1)\,$. Second,
$\alpha_{\scriptscriptstyle F_2\circ F_1}(x_1)=(F_2\circ
F_1)(x_1)=\alpha_{\scriptscriptstyle
F_2}(F_1(x_1))=(\alpha_{\scriptscriptstyle
F_2}\circ\alpha_{\scriptscriptstyle F_1})(x_1)\,$. Third,
$F_\alpha^\ast(1_2)\!=\!G_\alpha(1_2)\!=\!\alpha^\ast(1_2)\!=\!a_1$, and
$\alpha_{\scriptscriptstyle
F_\alpha}(x_1)\!=\!F_\alpha(x_1)\!=\!\alpha(x_1)$
for $x_1<a_1$.
\par\noindent (3) If $x_2\not={\bf1}$ then 
$g^u(x_2)=g(x_2)=g^p(x_2)=\alpha^\ast(x_2)=G_\alpha(x_2)\,$, whereas if
$x_2={\bf1}$ then $g^u(x_2)={\bf1}=G_\alpha(x_2)\,$. Hence
$g^u=G_\alpha$ so that $F=F_\alpha\,$. Finally, if
$x_2\in L_2$ then 
$g^p(x_2)=g(x_2)=g^u(x_2)=F^\ast(x_2)=\beta_{\scriptscriptstyle
F}(x_2)\,$,
so that $g^p=\beta_{\scriptscriptstyle F}$ and 
$\alpha=\alpha_{\scriptscriptstyle F}\,$.
\par\bigskip\par\noindent
\prf{5.1}
\par\noindent (1) Suppose that $f(\Sigma_{\scriptscriptstyle
L_1})\subseteq\Sigma_{\scriptscriptstyle L_2}\cup\{0_2\}\,$, and
$p_1\in\Sigma_{\scriptscriptstyle L_1}\,$. Then either $f(p_1)=0_2$ or
$f(p_1)$ is an atom. In either case there exists
$p_2\in\Sigma_{\scriptscriptstyle L_2}$ such that $f(p_1)<p_2$ and so
$p_1<g(p_2)\,$. Suppose that for each
$p_1\in\Sigma_{\scriptscriptstyle L_1}$ there exists
$p_2\in\Sigma_{\scriptscriptstyle L_2}$ such that $p_1<g(p_2)\,$. Then
$f(p_1)<p_2\,$, so that either $f(p_1)=p_2$ or $f(p_1)=0_2\,$.
\par\noindent (2) Let $\alpha(T_1A_1\setminus K_1)\subseteq
T_2\alpha(A_1\setminus K_1)\,$, and for $T_2A_2=A_2$ set
$A_1=K_1\cup\alpha^{-1}(A_2)\,$. Then 
$T_1A_1\!\subseteq\!K_1\cup\alpha^{-1}(\alpha(T_1A_1)\!\setminus\!K_1)
\!\subseteq\!K_1\!\cup\!\alpha^{-1}(T_2\alpha(A_1\!\setminus\!K_1))
\!\subseteq\!K_1\!\cup\!f^{-1}(A_2)\!=\!A_2$ since 
$A_1\!\setminus\!K_1\!=\!f^{-1}(A_2)$. Let
$T_1(K_1\cup\alpha^{-1}(A_2))=K_1\cup\alpha^{-1}(A_2)$ if
$T_2(A_2)\!=\!A_2$, and set $A_2\!=\!T_2\alpha(A_1\!\setminus\!K_1)$. Then 
$\alpha(T_1A_1\setminus K_1)\subseteq\alpha(\alpha^{-1}(A_2))\subseteq
A_2=T_2\alpha(A_1\setminus K_1)$ since $A_1\subseteq
K_1\cup\alpha^{-1}(\alpha(A_1\setminus K_1))\subseteq
K_1\cup\alpha^{-1}(T_2\alpha(A_1\setminus
K_1))=K_1\cup\alpha^{-1}(A_2)\,$.
\par\noindent (3) First,  
$(f_2\circ f_1)(\Sigma_{\scriptscriptstyle L_1})\subseteq
f_2(\Sigma_{\scriptscriptstyle
L_2}\cup\{0_2\})\subseteq\Sigma_{\scriptscriptstyle L_3}\cup\{0_3\}\,$.
Second, the kernel of a composition is given by
$K=K_1\cup\alpha_1^{-1}(K_2)\,$. Hence, if $A_3$ is closed then so is
$K\!\cup\!(\alpha_2\!\circ\!\alpha_1)^{-1}(A_3)
\!=\!(K_1\!\cup\!\alpha_1^{-1}(K_2))\!\cup\!\alpha_1^{-1}(\alpha_2^{-1}(A_3))
\!=\!K_1\!\cup\!\alpha_1^{-1}(K_2\!\cup\!\alpha_2^{-1}(A_3))$.
Third, 
$K_{\alpha_3\!\circ\!(\alpha_2\!\circ\!\alpha_1)}
\!=\!(K_1\!\cup\!\alpha_1^{-1}(K_2))\!\cup
\!(\alpha_2\!\circ\!\alpha_1)^{-1}(K_3)
\!=\!K_1\!\cup\!\alpha_1^{-1}(K_2\!\cup\!\alpha_2^{-1}(K_3))
\!=\!K_{(h\!\circ\!g)\!\circ\!f}\,$.
\par\bigskip\par\noindent
\prf{5.2}
\par\noindent (1) 
$A=\{\,p\in\Sigma\,|\,p\in A\,\}\subseteq\{\,p\in\Sigma\,|\,p<\bigvee
A\}=\{\,p\in\Sigma\,|\,p<\pi(A)\,\}=(i\circ\pi)(A)$ and
$(\pi\circ i)(a)=\bigvee i(a)=\bigvee\{\,p\in\Sigma\,|\,p<a\,\}=0\,$, so
that $\pi\dashv i\,$. Further, the closure
$i\circ\pi$ is simple, since
$(i\circ\pi)(\emptyset)=\{\,p\in\Sigma\,|\,p<\bigvee\emptyset=0\,\}=
\emptyset$
and
$(i\circ\pi)(\{p\})=\{\,q\in\Sigma\,|\,p<\bigvee\{p\}=p\,\}=\{p\}\,$.
\par\noindent (2) Note that $A\in{\rm P}(\Sigma)^{i\circ\pi}$ if and only
if 
$A=\{\,p\!\in\!\Sigma\,|\,p\!<\!a\,\}\,$. Let
$(i_2\!\circ\!\pi_2)(A_2)=A_2\,$. Then 
$K_1=\{\,p\in\Sigma_1\,|\,f(p_1)=0_2\,\}=\{\,p\in\Sigma_2\,|
\,f(p_1)=0_2\,{\&}\,f(p_1)<\bigvee
A_2\,\}$ and
$\alpha_f^{-1}(A_2)=\{\,p_1\in\Sigma_1\,|\,\alpha_f(p_1)\in
A_2\,\}=\{\,p_1\in\Sigma_1\,|\,f(p_1)\not=0\,{\&}\,f(p_1)<\bigvee
A_2\,\}\,$, so that
$K_1\!\cup\!\alpha_f^{-1}(A_2)
\!=\!\{p_1\!\in\!\Sigma_1\,|\,f(p_1)\!<\!\bigvee A_2\}
\!=\!\{p_1\!\in\!\Sigma_1\,|\,p_1\!<\!f^\ast(\bigvee A_2)\}$ is closed.
\par\noindent (3) First, if $f_\alpha(A_1)\subseteq A_2$ then 
$\alpha(A_1\setminus K_1)\subseteq T_2\alpha(A_1\setminus
K_1)=f_\alpha(A_1)\subseteq A_2$ so that
$A_1\!\setminus\!K_1\!\subseteq\!\alpha^{-1}(\alpha(A_1\!
\setminus\!K_1))\!\subseteq\!f^{-1}(A_2)$
and 
$A_1\!\subseteq\!K_1\!\cup\!(A_1\!\setminus\!K_1)\!\subseteq
\!K_1\!\cup\!\alpha^{-1}(A_2)\!=\!g_\alpha(A_2)\,$.
Second, if $A_1\subseteq g_\alpha(A_2)=K_1\cup\alpha^{-1}(A_2)$ then we
have
$A_1\setminus K_1\subseteq\alpha^{-1}(A_2)$ so that
$f_\alpha(A_1)=T_2\alpha(A_1\setminus K_1)\subseteq
T_2\alpha(\alpha^{-1}(A_2))\subseteq T_2(A_2)=A_2\,$. Third, let
$p_1\in\Sigma_1\,$. Then $p_1\notin K_1\Rightarrow
f_\alpha(\{p_1\})=T_2\alpha(\{p_1\}\setminus
K_1)=T_2f(\{p_1\})=T_2\{f(p_1)\}=\{f(p_1)\}\,$, whereas 
$p_1\in K_1\Rightarrow
f_\alpha(\{p_1\})=T_2\alpha(\{p_1\}\setminus
K_1)=T_2f(\emptyset)=T_2(\emptyset)=\emptyset\,$.
\par\noindent (4) 
${\cal L}:\alpha\!\mapsto\!f_\alpha$ and ${\cal C}:f\!\mapsto\!\alpha_f$ 
are functors, since
$K_{\alpha_2\!\circ\!\alpha_1}\!=\!K_{\alpha_1}\!
\cup\!\alpha_1^{-1}(K_{\alpha_2})$
so that
$g_{\alpha_2\!\circ\!\alpha_1}(A_3)
\!=\!K_{\alpha_2\!\circ\!\alpha_1}\!\cup\!(\alpha_2\!\circ\!\alpha_1)^{-1}(A_3)
\!=\!K_{\alpha_1}\!\cup\!\alpha_1^{-1}(K_{\alpha_2}\!\cup\!\alpha_2^{-1}(A_3))
\!=\!(g_{\alpha_1}\!\circ\!g_{\alpha_2})(A_3)$, 
and
$K_{\alpha_{f_2\circ
f_1}}\!=\!\{\,p_1\!\in\!\Sigma_1\,|\,(f_2\!\circ\!f_1)\!=\!0_3\,\}
\!=\!K_{\alpha_{f_1}}\!\cup\alpha_1^{-1}(K_{\alpha_{f_2}})
\!=\!K_{\alpha_{f_2}\circ\alpha_{f_1}}$
so that $\alpha_{f_2\!\circ\!f_1}\!=\!\alpha_{f_2}\!\circ\!\alpha_{f_1}$.
Let 
$\phi_{\scriptscriptstyle\Sigma}
\!:\!\Sigma\!\rightarrow\!{\cal CL}\Sigma
\!:\!p\!\mapsto\!\{p\}$ and 
$\psi_{\scriptscriptstyle L}
\!:\!L\!\rightarrow\!{\cal LC}L
\!:\!a\!\mapsto\!\{p\!\in\!\Sigma_{\scriptscriptstyle L}\,|\,p\!<\!a\}\,$. 
Then $\phi\,$, $\psi$ are natural, 
$({\cal CL}\alpha\!\circ\!\phi_{\scriptscriptstyle\Sigma_1})(p)
\!=\!\alpha_{f_\alpha}(\{p\})
\!=\!f_\alpha(\{p\})
\!=\!\{\alpha(p)\}
\!=\!(\phi_{\scriptscriptstyle\Sigma_2}\!\circ\!\alpha)(p)$
and
$({\cal LC}f\!\circ\!\psi_{\scriptscriptstyle L_1})(a)
\!=\!f_{\alpha_f}(\psi_{\scriptscriptstyle L_1}(a))
\!=\!\{p\!\in\!\Sigma_{\scriptscriptstyle L_1}\,|\,p\!<\!f(a)\}
\!=\!(\psi_{\scriptscriptstyle L_2}\!\circ\!f)(a)$, 
and form an equivalence,
${\cal
L}\phi_{\scriptscriptstyle\Sigma}(A)=f_{\phi_{\Sigma}}(A)=
\phi_{\scriptscriptstyle\Sigma}(TA)=
\phi_{\scriptscriptstyle\Sigma}(A)=\{\,\{p\}\,|\,\{p\}\subseteq
A\,\}=\psi_{\scriptscriptstyle{\cal L}\Sigma}(A)\,$.
\par\bigskip\par\noindent
\prf{5.3}
\par\noindent (1) First, the maps $S_\alpha$ and $P_\alpha$ are
isotone, since for 
$A_2\subseteq B_2$ and $A_1\subseteq B_1$ we have
$x_1\in S_\alpha(A_1)
\Rightarrow(\exists x_2\in A_2)\,\alpha(x_1)=x_2
\Rightarrow(\exists x_2\in B_2)\,\alpha(x_1)=x_2
\Rightarrow x_1\in S_\alpha(B_2)$ 
and
$x_2\!\in\!P_\alpha(A_2)
\Rightarrow(\exists x_1\!\in\!A_1)\,\alpha(x_1)\!=\!x_2
\Rightarrow(\exists x_1\!\in\!B_1)\,\alpha(x_1)\!=\!x_2
\Rightarrow x_2\!\in\!P_\alpha(B_1)\,$. 
Second,
$x_1\in(S_\alpha\circ P_\alpha)(A_1)\Leftrightarrow\alpha(x_1)\in
P_\alpha(A_1)\Leftrightarrow(\exists y_1\in
A_1)\,\alpha(x_1)=\alpha(y_1)\,$. In particular, 
$A_1\!\subseteq\!(S_\alpha\!\circ\!P_\alpha)(A_1)\,$, and if 
$S_\alpha\!\circ\!P_\alpha\!=\!{\rm id}$ or $\alpha$ is injective we 
have that
$\alpha(x_1)\!=\!\alpha(\,\overline{x}{}_1)
\!\Leftrightarrow\!(\exists y_1\!\in\!\{\,\overline{x}{}_1\})\,
\alpha(x_1)\!=\!\alpha(y_1)
\!\Leftrightarrow\!x_1\!\in\!(S_\alpha\!\circ\!P_\alpha)
(\{\,\overline{x}{}_1\})\!=\!\{\,\overline{x}{}_1\}\!\Leftrightarrow\!
x_1\!=\!\overline{x}{}_1$ and
$x_1\!\in\!(S_\alpha\!\circ\!P_\alpha)(A_1)\Leftrightarrow
(\exists y_1\!\in\!A_1)\,
[\alpha(x_1)\!=\!\alpha(y_1)\Leftrightarrow x_1\!=\!y_1]
\Leftrightarrow x_1\!\in\!A_1$. 
Third,
$x_2\in(P_\alpha\circ S_\alpha)(A_2)\Leftrightarrow(\exists x_1\in
S_\alpha(A_2))\,\alpha(x_1)=x_2\Leftrightarrow(\exists
x_1\in\Sigma_1)\,x_2=\alpha(x_1)\in A_2\,$.  In particular, 
$(P_\alpha\circ S_\alpha)(A_2)\subseteq A_2\,$, and if $P_\alpha\circ
S_\alpha={\rm id}$ or $\alpha$ is surjective we obtain 
$x_2\in\Sigma_2\Leftrightarrow x_2\in\{x_2\}=(P_\alpha\circ
S_\alpha)(\{x_2\})\Leftrightarrow(\exists
x_1\in\Sigma_1)\,x_2=\alpha(x_1)$ and
$x_2\in(P_\alpha\circ S_\alpha)(A_2)\Leftrightarrow(\exists
x_1\in\Sigma_1)\,x_2=\alpha(x_1)\in A_2\Leftrightarrow x_2\in A_2\,$.
\par\noindent (2) First, 
$p=p\wedge1=p\wedge(a\vee a')=(p\wedge a)\vee(p\wedge a')\,$, 
so that either $p\wedge a=p\Leftrightarrow p<a$ or 
$p\wedge a'=p\Leftrightarrow p<a'$. 
In particular $p\not<q'$ if and only if $p=q\,$, so that for
$A\subseteq\Sigma_{\scriptscriptstyle B}$ we obtain
$p<\bigvee A\Leftrightarrow p\not<(\bigvee A)'=\bigwedge
A'\Leftrightarrow(\exists q\in A)\,p\not<q'\Leftrightarrow p\in A\,$.
Hence
$(\rho\circ\mu)(a)=\bigvee\{\,p\in\Sigma_{\scriptscriptstyle
B}\,|\,p<a\,\}=a$ and
$(\mu\circ\rho)(A)=\{\,p\in\Sigma_{\scriptscriptstyle B}\,|\,p<\bigvee
A\,\}=A\,$. Second, let $f(\Sigma_{\scriptscriptstyle
B_1})\subseteq\Sigma_{\scriptscriptstyle B_2}\,$. Then $g(0_2)=0_1\,$,
since if $p_1<g(0_2)$ then $f(p_1)<0_2$ which is impossible by hypothesis,
so that $g(a_2)\wedge g(a_2')=g(a_2\wedge a_2')=g(0_2)=0_1\,$. Further,
for each $p_1\in\Sigma_{\scriptscriptstyle B_1}$ either
$f(p_1)<a_2$ and $p_1<g(a_1)<g(a_1)\vee g(a_1')$ or $f(p_1)<a_2'$ and
$p_1<g(a_2')<g(a_2)\vee g(a_2')\,$, so that $g(a_2)\vee g(a_2')=1_1\,$.
Hence $g(a_2')=g(a_2)'$.  Third, let $g(a_2')=g(a_2)'$. Then
$f(p_1)\!\not=\!0_2$ since we have that
$f(a_1)\!<\!0_2\Leftrightarrow
a_1\!<\!g(0_2)\!=\!g(1_2')\!=\!g(1_2)'\!=\!1_1'\!=\!0_1\,$. Further, if
$a_2<f(p_1)$ then either $p_1<g(a_2)\,$, so that $a_2<f(p_1)<a_2$ and
$a_2=f(p_1)\,$, or $p_1<g(a_2)'=g(a_2')\,$, so that $a_2<f(p_1)<a_2'$ and
$a_2=0_2\,$. Hence $f(p_1)$ is an atom. 
\par\bigskip\par\noindent
\prf{6.1}
\par\noindent (1) The maps are isotone$\,$: if $A\subseteq B$ then 
$J(A)=\bigvee A<\bigvee B=J(B)\,$, and if $a<b$ then
$\ell(a)=[0,a]\subseteq[0,b]=\ell(b)\,$.  Further, 
$(J\circ\ell)(a)=J(\,[0,a]\,)=\bigvee\,[0,a]=a\,$,  and
$A\subseteq[0,\bigvee A]=\ell(\,\bigvee A)=(\ell\circ J)(a)\,$. 
\par\noindent (2) Let $f\succ\theta\,$. Then
$f=f\circ{\rm id}_1=f\circ J_1\circ\ell_1=J_2\circ\theta\circ\ell_1\,$. 
Further, let
$g=J_1\circ\phi\circ\ell_2$ where $\theta\dashv\phi\,$. Then
$f\circ g=f\circ
J_1\circ\phi\circ\ell_2=J_2\circ\theta\circ
\phi\circ\ell_2<J_2\circ\ell_2={\rm
id}_2$ and
${\rm
id}_1=J_1\circ\ell_1<J_1\circ\phi\circ\theta\circ
\ell_1<J_1\circ\phi\circ\ell_2\circ
J_2\circ\theta\circ\ell_1=g\circ f\,$, so that $f\dashv g\,$.
\par\noindent (3) Suppose that $\theta$ is strongly isotone, and define 
$f:L_1\rightarrow L_2:a_1\mapsto\bigvee\theta(\{a_1\})\,$. Then we have 
$f(\bigvee A_1)=\bigvee\theta(\{\bigvee A_1\})=\bigvee\theta(A_1)\,$,
since $\bigvee\{\bigvee A_1\}=\bigvee A_1$ so that
$\{\bigvee A_1\}\equiv A_1\,$. Suppose that
$f\succ\theta$ and $A_1\ll B_1\,$. Then
$\bigvee A_1<\bigvee B_1\,$, so that 
$\bigvee\theta(A_1)=f(\bigvee A_1)<f(\bigvee B_1)=\bigvee\theta(B_1)$ and
$\theta(A_1)\ll\theta(B_1)\,$. 
\par\noindent (4) If $f_1\succ\theta_1\,$, $f_2\succ\theta_2$ then 
$f_2\circ f_1\circ J_1=f_2\circ
J_2\circ\theta_1=J_3\circ\theta_2\circ\theta_1\,$, and $f_2\circ
f_1\succ\theta_2\circ\theta_1\,$. If 
$f_\alpha\succ\theta_\alpha$ then 
$(\vee_\alpha f_\alpha)\circ J_1=\vee_\alpha(f_\alpha\circ
J_1)=\vee_\alpha(J_2\circ\theta_\alpha)=J_2\circ
(\cup_\alpha\theta_\alpha)\,$,
and $\vee_\alpha f_\alpha\succ\cup_\alpha\theta_\alpha\,$.
\par\bigskip\par\noindent
\prf{6.2}
\par\noindent (1) First, $f\in J({\bf2},L)$ iff 
$f=f_a:[0\mapsto0;\,1\mapsto a]\,$, whereas 
$\theta\in J(P_0({\bf2}),P_0(L))$ iff 
$\theta=\theta_{\scriptscriptstyle
A}:[\emptyset\mapsto\emptyset;\,\{1\}\mapsto A]\,$. Second, 
$P_{f_a}=\theta_{\{a\}}$ so that
$\cup_{a\in A}P_{f_a}=\theta_{\scriptscriptstyle A}\,$. Third, 
$f_a\!\succ\!\theta_{\scriptscriptstyle A}
\Leftrightarrow f_a(\bigvee X)\!=\!\bigvee\theta_{\scriptscriptstyle A}(X)
\Leftrightarrow
a\!=\!f(1)\!=\!f(\bigvee\{1\})\!=\!\bigvee\theta(\{1\})\!=\!\bigvee A\,$.
Hence $P\!S({\bf2},L)=L$ whereas 
$B\!S({\bf2},L)=T\!S({\bf2},L)=F\!S({\bf2},L)=P_0(L)\,$.
\par\noindent (2) First, $f\in J(L,{\bf2})$ iff 
$f=f_a:[x\mapsto0\,(x<a)\,;\ 1\,(x\not<a)]\,$, whereas we have that
$\theta\in J(P_0(L),P_0({\bf2}))$ iff
$\theta=\theta_{\scriptscriptstyle A}:[X\mapsto\emptyset\,(X\subseteq
A)\,;\ \{1\}\,(X\not\subseteq A)]\,$. Second, 
$P_{f_a}(\{x\})=[\emptyset\,(x<a)\,;\ \{1\}\,(x\not<a)]$ and so
$P_{f_a}=\theta_{[0,a]}\,$. Third,  
$f_a\succ\theta_{\scriptscriptstyle A}$ iff
$\bigvee X<a\Leftrightarrow f_a(\bigvee
X)=0\Leftrightarrow\bigvee\theta_{\scriptscriptstyle
A}(X)=0\Leftrightarrow X\subseteq A\,$, iff $A=[0,a]$ and
$\theta_{\scriptscriptstyle A}=P_{f_a}\,$. Hence 
$P\!S(L,{\bf2})=B\!S(L,{\bf2})=T\!S(L,{\bf2})=L$ whereas 
$F\!S(L,{\bf2})=P_0(L)\,$. 
\par\noindent (3) We must give an example for which the inclusion
$\hbox{\underbar{BStruct}}\hookrightarrow\hbox{\underbar{TStruct}}$ is
strict. Let
$\theta\!:\!P_0(L)\!\rightarrow\!P_0(L)\!:\!A\!\mapsto\![A\,(1\!\notin\!A);\,
\{a\}\!\cup\!A\,(1\!\in\!A)]$. Then $\theta$ preserves unions,
$\theta(\cup_\alpha\!A_\alpha)\!=\!\{a\}\!\cup\!
A\!\Leftrightarrow\!1\!\in\!\cup_\alpha\!
A_\alpha\!\Leftrightarrow\!(\exists\alpha)\,[1\!\in
\!A_\alpha\!\Leftrightarrow\!\theta(A_\alpha)\!=\!\{a\}\!\cup\!
A]\!\Leftrightarrow\!\cup_\alpha\theta(A_\alpha)\!=\!\{a\}\!\cup\!A$, and ${\rm
id}_{\scriptscriptstyle L}\succ\theta\,$, since if $1\notin A$ we have
${\rm id}_{\scriptscriptstyle L}(\bigvee A)=\bigvee A=\bigvee\theta(A)$
whereas if $1\in A$ we have 
${\rm id}_{\scriptscriptstyle L}(\bigvee A)=\bigvee
A=1=a\vee1=a\vee(\bigvee A)=\bigvee(\{a\}\cup A)=\bigvee\theta(A)$.
However, if there exists $1\not=b\not<a$ then we cannot express $\theta$
as a union of power maps. Indeed, suppose that $\theta=\cup_\alpha
P_{f_\alpha}$ so that 
$\theta(\{x\})=\cup_\alpha P_{f_\alpha}(\{x\})=\{f_\alpha(x)\}$.  Then 
$\{f_\alpha(x)\}=\theta(\{x\})=\{x\}$ for $x\not=1$ whereas
$\{f_\alpha(1)\}=\theta(\{1\})=\{a,1\}$. Hence we would have
$\theta=P_{f_1}\cup P_{f_a}$ where $f_1:x\mapsto x$ and $f_a:x\mapsto
x\,(x\not=1)\,;\ a\,(x=1)$, the latter not being isotone since $b<1$
whereas $f_a(b)=b\not<a=f_a(1)$.

\end{document}